\documentclass[aps,prl,twocolumn,superscriptaddress,longbibliography]{revtex4-1}

\usepackage{times}
\usepackage{graphicx, amsmath, verbatim, dsfont, amsfonts, color}
\usepackage{bbm, bm, latexsym, amssymb}
\usepackage{graphicx}
\usepackage{epstopdf}
\usepackage{dcolumn}
\usepackage{bm}
\usepackage{gensymb}
\usepackage{upgreek}
\usepackage{hyperref}
\usepackage{graphicx}
\usepackage{epstopdf}
\usepackage{dcolumn}
\usepackage{bm}
\usepackage{gensymb}
\usepackage{upgreek}
\usepackage{booktabs}
\usepackage{hyperref}

\usepackage{amssymb,mathtools}
\usepackage{color}
\usepackage{float}

\usepackage{tikz,xcolor,hyperref}

\begin{document}
\author{Tomasz~Dietl}
\email{dietl@MagTop.ifpan.edu.pl}
\affiliation{International Research Centre MagTop, Institute of Physics, Polish Academy of Sciences, Aleja Lotnikow 32/46, PL-02668 Warsaw, Poland}
\affiliation{WPI Advanced Institute for Materials Research, Tohoku University, 2-1-1 Katahira, Aoba-ku, Sendai 980-8577, Japan}

\title{Effects of charge dopants in quantum spin Hall materials}

\begin{abstract}
Semiconductors' sensitivity to electrostatic gating and doping accounts for their widespread use in information communication and new energy technologies. It is demonstrated quantitatively and with no adjustable parameters that the presence of paramagnetic acceptor dopants elucidates a variety of hitherto puzzling properties  of two-dimensional topological semiconductors at the topological phase transition and in the regime of the quantum spin Hall effect. The concepts of resonant states, charge correlation, Coulomb gap, exchange interaction between conducting electrons and holes localized on acceptors, strong coupling limit of the Kondo effect, and bound magnetic polaron explain  a short topological protection length, high hole mobilities compared with electron mobilities, and different temperature dependence of the spin Hall resistance in HgTe and (Hg,Mn)Te quantum wells.
\end{abstract}
\maketitle

{\em{Introduction}}--Quantized Hall resistance is a hallmark of two-dimensional (2D) topological electronic systems \cite{Klitzing:2020_NRP}. The integer quantum Hall effect's high-precision quantization is behind a new definition  of units \cite{Rigosi:2019_SST}, whereas other quantum Hall phenomena lead to many far-reaching developments \cite{Klitzing:2020_NRP}. Surprisingly, however, although the quantum spin Hall effect (QSHE) has been known for more than a decade \cite{Kane:2005_PRL,Bernevig:2006_S,Konig:2007_S}, experimental resistance magnitudes attain the expected value $h/2e^2$  only in mesoscopic samples, such as micron-size HgTe-based quantum wells (QWs) \cite{Bendias:2018_NL,Shamim:2021_NC} and sub-100-nm atomically thin 1T'-WTe$_2$ 2D monolayers \cite{Fei:2017_NP,Wu:2018_S}. Moreover, although several theoretical models have been proposed \cite{Hsu:2021_SST,Yevtushenko:2022_NJP}, a short experimentally found protection length has usually been assigned \cite{Bendias:2018_NL,Shamim:2021_NC,Fei:2017_NP,Wu:2018_S} to unidentified charge puddles that trap edge carriers and within which spin-flip, allowing for scattering between helical edges, occurs \cite{Vayrynen:2014_PRB}.

We claim here that the challenging properties of QSHE semiconductors result from the presence of native acceptors in these materials. Quantitative
agreement between experimental and theoretical values of the topological protection lengths supports this claim. The starting point for this work is a quantitative theory of acceptor states in HgTe QWs, which provides positions of acceptor levels with respect to bands and topological edge states as a function of the QW thickness. With this information, we contend that the acceptor density is determined by the gate voltage range in which edge states carry the electric current
\cite{Konig:2007_S,Grabecki:2013_PRB,Bendias:2018_NL,Lunczer:2019_PRL,Shamim:2021_NC,Fei:2017_NP,Wu:2018_S}.
 Furthermore, considering charge correlation and  Coulomb-gap effects \cite{Wilamowski:1990_SSC}, the acceptor scenario explains why at the 2D topological phase transition, the mobility of holes is significantly greater than that of electrons \cite{Shamim:2020_SA,Yahniuk:2021_arXiv}, as well as elucidates the origin of high-frequency conductivity \cite{Dartiailh:2020_PRL} and gating hystereses \cite{Lunczer:2019_PRL}.  As a next step, a theory of exchange coupling between electrons and acceptor holes \cite{Sliwa:2008_PRB} is employed to demonstrate that, in topological materials, the interaction between edge electrons with acceptor holes reaches the strong coupling limit of the Kondo effect, where the spin dephasing rate assumes, up to a material-specific logarithmic correction, a universal behavior discussed in the context of magnetic impurities \cite{Maciejko:2009_PRL,Tanaka:2011_PRL,Micklitz:2006_PRL}. The central result of this work is that, in this limit, the topological protection length $L_{\text{p}}$ in the Ohmic conductivity regime is given by a product of the inverse of one-dimensional (1D) acceptor hole density in the edge region and the anisotropy of exchange coupling to hole spins. This finding elucidates the magnitude of $L_{\text{p}}$ in HgTe QWs and WTe$_2$ 2D monolayers. Finally, we demonstrate that the formation of acceptor bound magnetic polarons explains a difference in carrier mobilities and the temperature dependence of the edge resistivity of topological HgTe and Hg$_{1-x}$Mn$_x$Te QWs \cite{Shamim:2021_NC}. The result presented here are supported and extended in the companion paper \cite{Dietl:2023_PRB}.

{\em{Acceptor levels}}--Electrically active point centers, together with planar and linear defects, such as dislocations,  account for differences between devices fabricated to be similar. However, steady and impressive progress in the quality of MBE-grown modulation-doped III-V \cite{Umansky:2009_JCG,Chung:2021_NM} and II-VI \cite{Tsukazaki:2010_NM,Piot:2010_PRB,Betthausen:2014_PRB} heterostructures has been achieved by increasing pumping efficiency and improving chemical purity of constituting elements, which points to the dominant role of background ionized donor and acceptor impurities in epitaxial structures of those compound semiconductors. Similarly, native acceptors in bulk compound semiconductors have been frequently assigned to metal vacancies giving double acceptors ($Z = -2$) in II-VI materials, but the case of ZnTe and CdTe indicates that residual charged impurities, such as Cu ($Z = - 1$), are involved \cite{Pautrat:1985_JCG}.

To describe charge dopant states in topological QWs, the Kohn-Luttinger effective mass theory developed for acceptors in GaAs and HgTe QWs taking into account four $\Gamma_8$ valence bands (including spin) \cite{Fraizzoli:1991_PRB,Kozlov:2019_S} is extended in this Letter to the case, in which also $\Gamma_6$ and $\Gamma_7$ bands are relevant \cite{Novik:2005_PRB}. In the companion paper \cite{Dietl:2023_PRB}, we present an explicit form of the wave functions that diagonalize the eight bands' QW Hamiltonians without and with a charged impurity, as well as examine the validity range of the axial approximation, employed routinely for acceptors in zinc-blende QWs \cite{Fraizzoli:1991_PRB,Kozlov:2019_S}. Within that approximation, the eigenfunctions are found to be labeled by eigenvalues of $F_z =j_z + l_z$, where $j_z$ and $l_z$ denote the components perpendicular to the QW plane of the angular momenta corresponding to the Kohn-Luttinger amplitudes and the associated envelope functions, respectively, confirming that $F_z$ commutes with axial eight bands' Hamiltonians \cite{Sercel:1990_PRB}. The resultant wave functions are mainly composed of  either $p_{\pm 1/2}$ and $s_{\pm 1/2}$ ($j_z = \pm 1/2$) or $p_{\pm 3/2}$ ($j_z = \pm 3/2$)  Kohn-Luttinger amplitudes, respectively, where $s_{sz}$ and $p_{jz}$  transform under the point group operations like $s$ and $p$ atomic orbitals. The corresponding binding energies of the ground-state Kramers doublets are denoted $E_{1/2}$ or $E_{3/2}$, and are usually referred to as light and heavy hole acceptors, respectively.

\begin{figure}[tb]
\hspace*{-1.3cm}
	\includegraphics[width=1.25\columnwidth, angle=0]{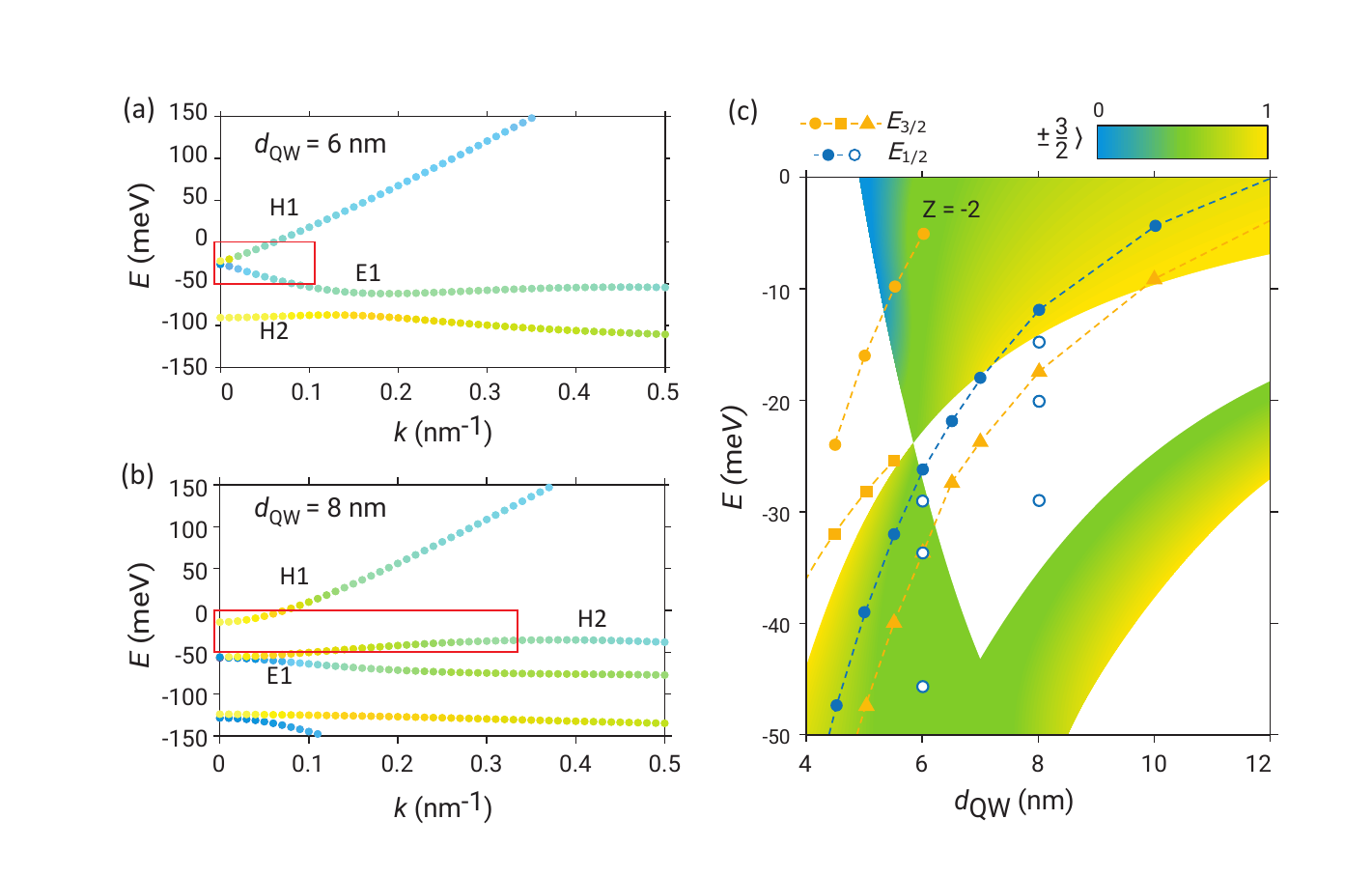}
	\caption{Band structure and positions of acceptor levels in HgTe QWs of different thicknesses computed with band structure parameters given in Ref.\,\onlinecite{Novik:2005_PRB}. (a, b) Band energies $E$ {\em vs.} wavevector $k$ for unstrained QW thickness of 6 and 8\,nm sandwiched between Hg$_{0.3}$Cd$_{0.7}$Te barriers. Red rectangles depict the band region displayed in (c). (c) Band edges and acceptor levels (symbols connected by dashed lines {\em vs.} QW thickness $d_{\text{QW}}$. Except for the orange circles computed for the doubly ionized acceptor ($Z = -2$), other symbols represent the single acceptor ($Z = -1$). The orange symbols ($E_{3/2}$)  correspond to acceptors associated with the valence band around $k = 0$; the blue symbols ($E_{1/2}$) with  valence band side maxima visible in (a) and (b). Full symbols represent the acceptors residing in the QW center; the open symbols represent acceptors at the distances $d_{\text{QW}}/4$, $d_{\text{QW}}/2$, and $3d_{\text{QW}}/2$ of the QW center. Colors represent the participation of the $p_{\pm3/2}$ Kohn--Luttinger amplitude in the wave functions. The discontinuity in the orbital content occurring at $k = 0$ and $E_{\text{g}} \rightarrow 0$ [see (a)], is blurred in (c) by contributions with $k\ne 0$.}
\end{figure}

Figure 1 depicts energies of relevant QW bands and acceptor ground-state levels for a range of the HgTe QW widths $d_{\text{QW}}$ with colors representing a fraction of the $p_{\pm 3/2}$ amplitude in the carrier wave function. Three distinct areas are observed in Fig.\,1(c): (i) normal band ordering (cation $s$ states above anion $p$ states) at small $d_{\text{QW}}$ values; (ii) the range of the topological phase transition centered around the bandgap $E_{\text{g}} = 0$ and $d_{\text{c}} \approx 5.8$\,nm; (iii) the topological region $d_{\text{w}} > d_{\text{c}}$, where the band ordering is inverted, resulting in 1D topological gapless edge states \cite{Kane:2005_PRL,Konig:2007_S} to be discussed later. Such a band diagram is generic for this class of 2D topological systems, however  the value of $d_{\text{c}}$ depends on strain (set to zero here) and Cd or Mn content in the barriers and well \cite{Shamim:2020_SA,Dietl:2023_PRB}.

We note that the binding energies of the doubly ionized acceptors $E^{(2-/-)}$ are irrelevant for the low-energy physics. In contrast, $E^{(-/0)}$ levels, residing near band edges or in the gap, are essential.  They originate from either single acceptors ($Z = -1$) or singly ionized double acceptors that, in the mean-field approach, have the same binding energy as single acceptors. As seen in Fig.\,1(c), in the regions of interest here ($d_{\text{QW}}\approx d_{\text{c}}$
and $d_{\text{QW}}> d_{\text{c}}$), the ground state corresponds to the level $E_{1/2}$ associated with the side maximum of the valence band visible in Figs.\,1(a) and 1(b). Notably, the acceptor levels form a band, as the hole binding energy depends on the location of the parent acceptor impurity with respect to the QW center, as shown in Fig.\,1(c).

Within this model, the range of gate voltage corresponding to sweeping over the bandgap $E_{\text{g}}$ at $d_{\text{QW}} > d_{\text{c}}$ directly provides the 2D areal density of relevant acceptors $N_{\text{a}}$,  with the experimental data implying $N_{\text{a}} \approx 10^{11}$\,cm$^{-2}$ for HgTe QWs \cite{Bendias:2018_NL,Yahniuk:2021_arXiv}, the value consistent with the areal hole concentration in undoped QWs \cite{Konig:2013_PRX}, and $N_{\text{a}} \approx 10^{13}$\,cm$^{-2}$ for WTe$_2$ \cite{Fei:2017_NP}. The $N_{\text{a}}$ for HgTe QWs corresponds to the three dimensional (3D) concentration of the order of  $N_{\text{A}} =3\cdot10^{16}$\,cm$^{-3}$, a typical magnitude for bulk HgTe \cite{Szlenk:1979_pssb_b} and Hg$_{1-x}$Mn$_x$Te \cite{Sawicki:1983_Pr}. For such a concentration, the holes are localized, as for the evaluated Bohr radius of 5\,nm, the Mott critical concentration is $1.4\cdot10^{17}$\,cm$^{-3}$. Next, we demonstrate
 that the presence of acceptors explains several hitherto puzzling properties of 2D topological insulators.

\begin{figure}[tb]
\hspace*{-1.2cm}
	\includegraphics[width=1.2\columnwidth, angle=0]{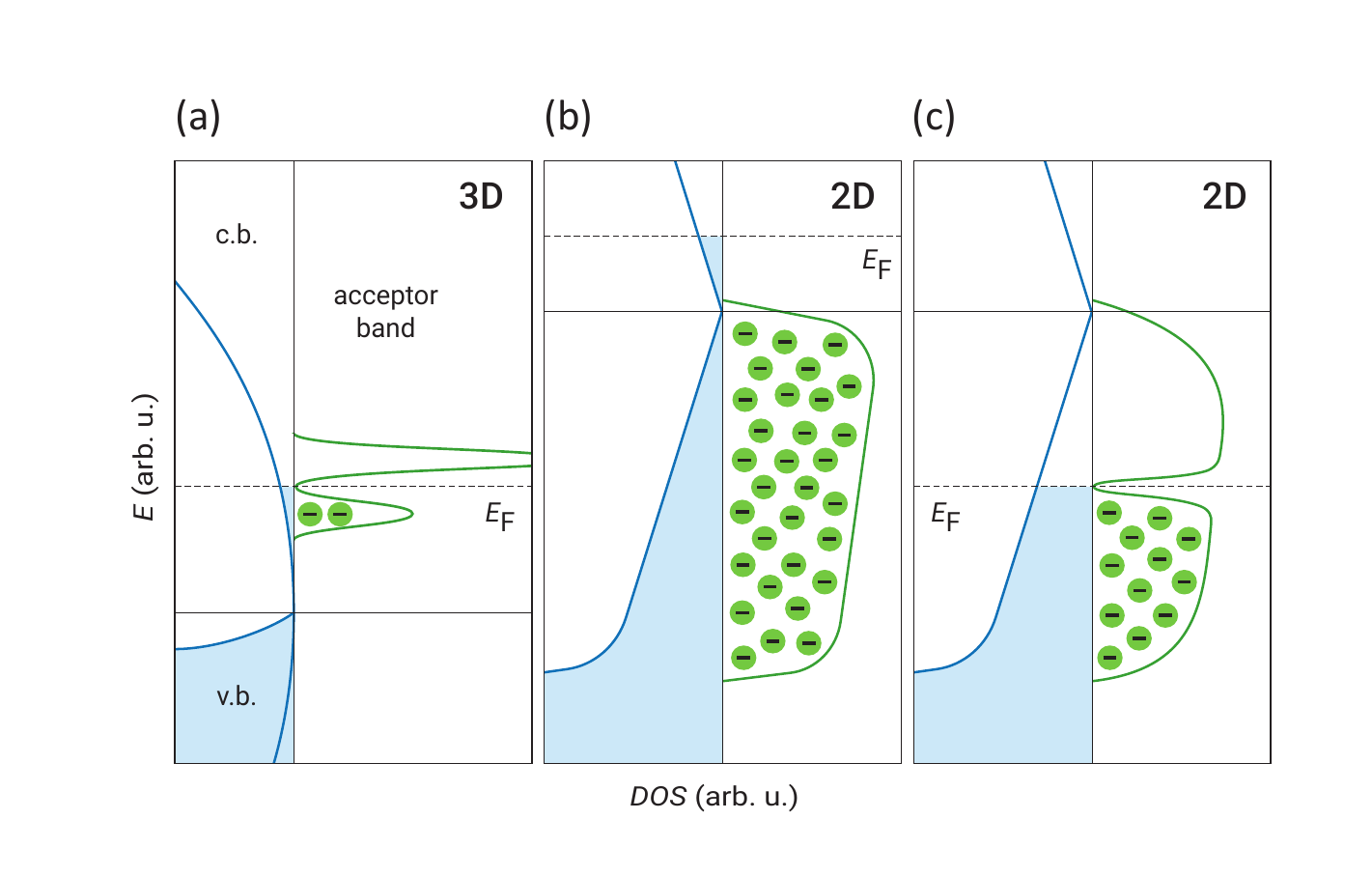}
	\caption{ Schematic picture of carrier and acceptor bands at the topological phase transition ($E_{\text{g}} =0$). (a) Bulk 3D case with the Fermi energy pinned in the conduction band (c.\,b.) by acceptors negatively charged below the Fermi level. Coulomb gap at $E_{\text{F}}$ is also shown. (b,c) The same for the 2D case and two positions of the Fermi level. The acceptor band is wide as the binding energy depends on the acceptor location with respect to the QW center.}
\end{figure}

{\em{Region of topological phase transition}}--One of the rather surprising facts is that low-temperature electron mobility $\mu_e$ in modulation donor-doped HgTe QWs $d_{\text{QW}}$ barely reaches  $0.4\cdot10^6$\,cm$^2$/Vs \cite{Bendias:2018_NL},  whereas $\mu_e$ in bulk HgTe as well as in Hg$_{1-x}$Cd$_x$Te and Hg$_{1-x}$Mn$_x$Te near the 3D topological transition approaches or exceeds $1\cdot10^6$\,cm$^2$/Vs \cite{Szlenk:1979_pssb_b,Dubowski:1981_JPCS,Sawicki:1983_Pr} with the onset of the Shubnikov de Haas oscillations at 10\,mT \cite{Sawicki:1983_Pr}. Even more surprisingly, in the vicinity of the topological phase transition in 2D QWs, the hole mobility $\mu_h$ is larger than $\mu_e$ \cite{Shamim:2020_SA,Yahniuk:2021_arXiv},
reaching $\mu_h = 0.9\cdot10^6$\,cm$^2$/Vs, for which the integer quantum Hall effect (QHE) plateau is resolved in 50\,mT in Hg$_{0.976}$Mn$_{0.024}$Te \cite{Shamim:2020_SA}, which is relevant for the QHE metrology \cite{Yahniuk:2021_arXiv}. In addition, the QW hole concentration evaluated from the Hall effect, is significantly smaller than the charge density generated by the gate voltage \cite{Shamim:2020_SA,Yahniuk:2021_arXiv}.

Figure 2 elucidates those findings using information obtained from Fig.\,1. In the 3D bulk case (Fig.\,2(a)), as previously discussed in detail \cite{Wilamowski:1990_SSC}, the acceptor band resides in the conduction band. In addition, due to a small electron mass value, we are on the metallic side of the Anderson--Mott transition so that donors do not bind electrons at any position of the Fermi energy $E_{\text{F}}$. Now, if the donor concentration $N_{\text{D}} \ll N_{\text{A}}$, most of the acceptors are neutral. Furthermore, under these conditions, to reduce the Coulomb energy, only acceptors in close vicinity to donors are ionized. The resulting dipole formation substantially reduces the electron scattering rate. Furthermore, the presence of the Efros--Shklovskii Coulomb gap precludes resonant scattering. By  fine hydrostatic pressure tuning of the band structure toward the 3D topological transition at $E_{\text{g}} = 0$,  $\mu_e = 20\cdot10^6$\,cm$^2$/Vs was registered in Hg$_{0.94}$Mn$_{0.06}$Te  at 2\,K \cite{Sawicki:1983_Pr}.

The situation is entirely different at the topological phase transition in the 2D case. As shown in Fig.\,2(b), for the Fermi level in the conduction band, obtained through modulation donor doping, all acceptors are ionized, explaining the low electron mobility. In contrast, in the hole transport regime [Fig.\,2(c)], achieved by gating-induced discharging of acceptors, the aforementioned charge correlation occurs, which along with the small effective mass of holes in the Dirac cone and the formation of the Coulomb gap $E_{\text{C}}$, results in high hole mobilities at $k_{\text{B}}T< E_{\text{C}} \approx 0.5$\,meV \cite{Dietl:2023_PRB}. However, with a growth of hole density, the hole effective mass increases [see, Fig.\,1(a)] and the hole mobility tends to diminish \cite{Shamim:2020_SA,Yahniuk:2021_arXiv}. Interestingly, higher carrier mobilities were observed in Mn-containing samples \cite{Shamim:2020_SA,Sawicki:1983_Pr} compared to the HgTe case. We note  that  $E_{\text{C}}$ is enlarged by the acceptor bound magnetic polaron (BMP) energy $E_{\text{p}}$, where for $x_{\text{Mn}} = 0.02$, $E_{\text{p}} > 0.3$\,meV  at $T <2$\,K \cite{Dietl:2023_PRB}. Further systematic experimental investigations would help verify the resonant BMP model proposed here. The presence of the Coulomb gap  explains also a large thermal stability of the QSHE in WTe$_2$ \cite{Wu:2018_S,Dietl:2023_PRB}.

{\em{Edge transport range}}--Having elucidated the role of acceptors in the region of the topological phase transition we focus on the region $d_{\text{QW}} > d_{\text{c}}$ (Fig.\,2(c)). Here, the Coulomb gap diminishes d.\,c.\,hoping conductivity.  However, since there is no Coulomb gap for electron--hole excitations, the presence of the acceptor band explains the origin of puzzling gap states detected by high-frequency conductivity \cite{Dartiailh:2020_PRL}. Moreover, under these conditions, one can anticipate the appearance of the exchange interaction ${\cal{H}}_{eh}$ between spins of electrons  in the topological edge states, ${\vec{s}}$, and paramagnetic acceptor holes, ${\vec{j}}$.

To reveal the striking consequences of this suggestion, it worth recalling that a long-range component of this coupling originates from the third order perturbation theory (second in \emph{kp} and first in the Coulomb interaction), for which the exchange energy ${\cal{J}}_{eh} \propto 1/E_{eh}^2$, where $E_{eh}$ represents the electron--hole energy distance \cite{Bir:1974_B}.
According to the theory \cite{Sliwa:2008_PRB}, which is quantitatively verified for the interaction between photoelectrons  at the bottom of the conduction and holes on Mn acceptors in GaAs \cite{Myers:2005_PRL}, ${\cal{H}}_{eh}$ assumes a scalar (Heisenberg) form, ${\cal{H}}_{eh} = -{\cal{J}}_{eh}{\vec{s}}\cdot{\vec{j}}$, where $j =3/2$ and ${\cal{J}}_{eh} = -0.23$\,eV \cite{Sliwa:2008_PRB}.
When $E_{eh} = 1.4$\,eV in GaAs:Mn, the lower bound of $E_{eh}$ is as small as $E_{\text{C}} \approx 0.3$\,meV for the topological edge electrons and acceptor holes.
Hence,  the antiferromagnetic ${\cal{J}}_{eh}$ is the largest relevant energy, and despite  a small DOS magnitude at $E_{\text{F}}$ in the 1D channels, drives the system to a strong coupling limit of the Kondo effect \cite{Dietl:2023_PRB}, specified in QWs by a wide distribution of Kondo temperatures $T_{\text{K}}$. For the parameter values specifying HgTe QWs, i.e., the Fermi velocity $v_{\text{F}} = 4\cdot10^5$\,m/s and the penetration length of the edge electron wave function into the QW, $b = 5$\,nm, a broad distribution of $T_{\text{K}}$ values up to 100\,K is expected \cite{Dietl:2023_PRB}. Importantly, for areal hole density $N_h = 0.5\cdot10^{11}$\,cm$^{-2}$, the number of edge electrons per unit length for $E_{\text{F}}$ in the gap center $n_e = E_{\text{g}}/2\pi\hbar v_{\text{F}} = 12/\mu$m is greater than the number of acceptor holes in the edge region, $n_h = N_hb = 3/\mu$m.
In the case of double acceptors, if the Hund's rule is obeyed, gating changes the value of holes' spin rather than $n_h$.

Thus, we can quantitatively verify numerous theoretical studies on the Kondo effect in QSHE materials \cite{Maciejko:2009_PRL,Tanaka:2011_PRL,Altshuler:2013_PRL,Eriksson:2013_PRB} and on the role of exchange anisotropy that allows for net backscattering of edge electrons \cite{Tanaka:2011_PRL,Altshuler:2013_PRL,Kimme:2016_PRB}. It worth noting in this context that for transition metal impurities such as Mn, $T_{\text{K}} \ll 1$\,mK in HgTe QWs and the exchange anisotropy vanishes if the transition metal is an orbital singlet state \cite{Dietl:2023_PRB}.

\begin{figure}[tb]
\hspace*{-0.8cm}
	\includegraphics[width=1.15\columnwidth, angle=0]{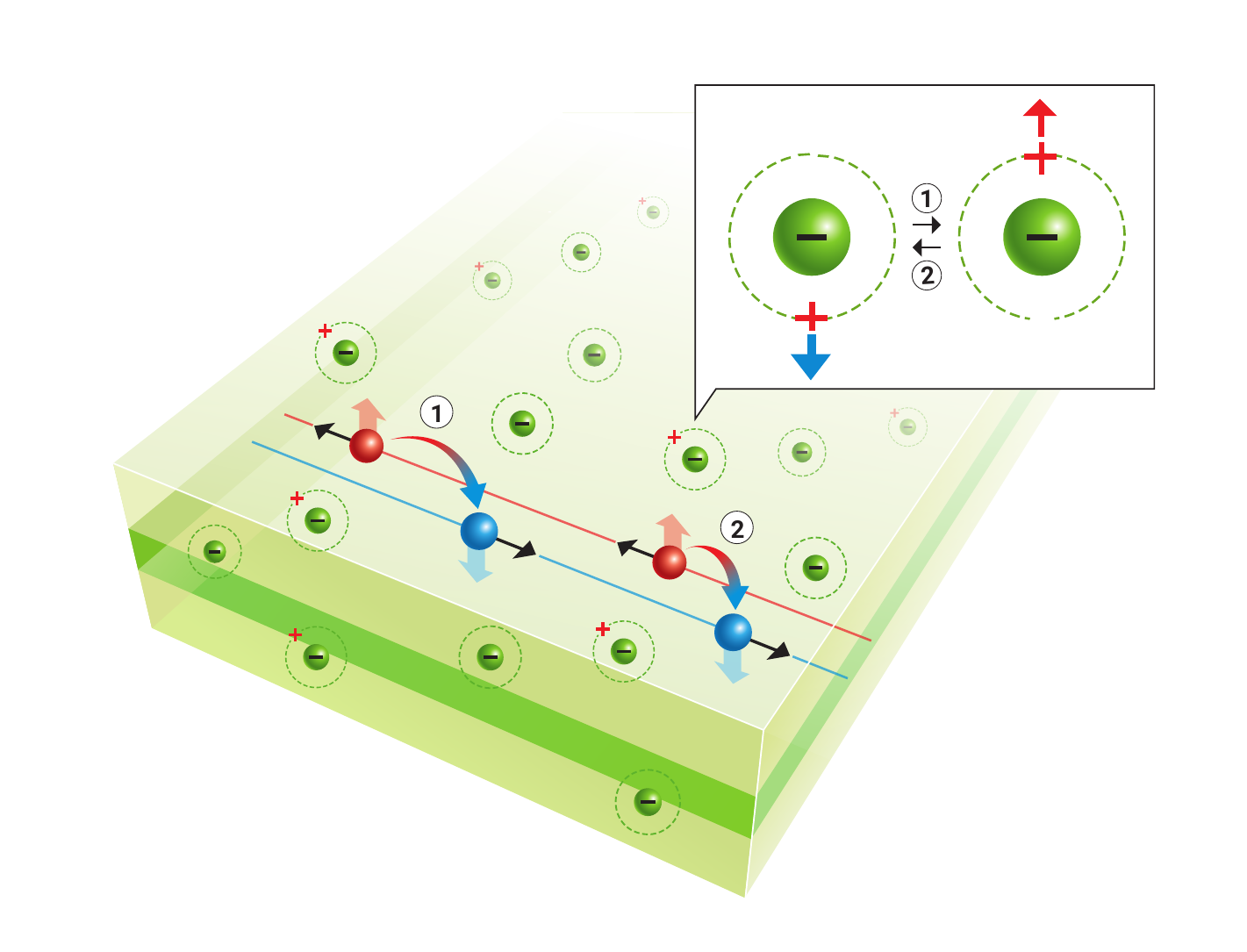}
	\caption{Destructive role of charge dopants in the quantum spin Hall effect. If axial symmetry is maintained (${\cal{J}}_x = {\cal{J}}_y), {\cal{J}}_{yz} = 0, D_x = 0$) only spin-flop ($\Uparrow\downarrow\, \leftrightarrows\, \Downarrow\uparrow$)
transitions occur (case 1 in the figure), so that edge current is conserved in the spin--momentum locking situation. Note that arrows up and down refer to time reversal partners rather than to spin up and down. However, if exchange interaction is anisotropic, $\Uparrow\uparrow\, \leftrightarrows\, \Downarrow\downarrow$
transitions that violate pseudospin conservation are also allowed (case 2), leading to net backscattering after a chain of spin-dependent interactions of electrons with an acceptor. For realistic concentrations of charge dopants,
backscattering is efficient in the strong coupling limit of the Kondo effect.}
\end{figure}

In general, the exchange Hamiltonian between pseudospins of edge electrons $s = 1/2$ and acceptor holes $j = 1/2$  assumes a form, ${\cal{H}}_{eh} = -\sum_{\alpha,\beta}s_{\alpha}{\cal{J}}^{(\alpha,\beta)}j_{\beta}$, where ${\cal{J}}^{(\alpha,\beta)}$ is a real tensor, whereas $\alpha$ and $\beta$ refer to the vector components $x,y,z$. It is convenient to introduce the notation ${\cal{J}}_{\alpha} ={\cal{J}}^{(\alpha,\alpha)}$ for $\alpha = \beta$ and if $\alpha \ne \beta$, ${\cal{J}}_{\alpha,\beta}^{(s)} =({\cal{J}}^{(\alpha,\beta)}+ {\cal{J}}^{(\beta,\alpha)})/2$ and $D_{\alpha} = \sum_{\beta,\gamma}\epsilon_{\alpha\beta\gamma}{\cal{J}}^{(\beta,\gamma)}/2$, where $D_{\alpha}$ are vector components
 of the Dzyaloshinskii-Moriya contribution and $\epsilon_{\alpha\beta\gamma}$ is the antisymmetric Levi-Civita tensor. If axial symmetry is maintained, the group theory implies ${\cal{J}}_x = {\cal{J}}_y, {\cal{J}}_{yz} = 0, D_x = 0$. In such a situation,  only spin-flop ($\Uparrow\downarrow\, \leftrightarrows\, \Downarrow\uparrow$) are allowed, which precludes net backscattering in the spin-momentum locking case, as sketched in Fig.~3 \cite{Tanaka:2011_PRL,Altshuler:2013_PRL,Kimme:2016_PRB}. Since, however, the edge breaks the axial symmetry and a random distribution of holes breaks the inversion symmetry, there appear anisotropic contributions $ {\cal{J}}_{\text{an}}$ of the form ${\cal{J}}_{x} -{\cal{J}}_{y}$, ${\cal{J}}_{yz}$, and $D_x$. The presence of such spin non-conserving terms  ensures the leak of electron angular momentum to crystal orbital momentum in a chain of scattering events and, thus, leads to net  backscattering of edge electrons \cite{Tanaka:2011_PRL,Altshuler:2013_PRL,Kimme:2016_PRB}. The resulting backscattering rate, compared to the conventional spin dephasing rate $\gamma_{\text{s}}$, is reduced by a factor $r$, so that $\gamma_{\text{b}} = r\gamma_s$, where $r =  [2{\cal{J}}_{\text{an}}/(({\cal{J}}_x + {\cal{J}}_y)]^2$ \cite{Tanaka:2011_PRL,Kimme:2016_PRB}.

 Using $\gamma_{\text{s}}$ determined by Wilson's numerical renormalization group approach for 1D systems in the Kondo regime \cite{Micklitz:2006_PRL}, and noting that  the topological protection length $L_{\text{p}} = v_{\text{F}}/\gamma_b$, we arrive to
 the main result of this Letter,
\begin{equation}
 L_{\text{p}}^{-1} = \sum_ir^{(i)}f(T/T_{\text{K}}^{(i)})/L_x,
 \label{eq:Lp}
\end{equation}
where the summation is over all QW holes bound to acceptors for a given gate voltage $V_{\text{g}}$. The function $F(x) = 1$ for $x=1$, it decays to zero for $x \rightarrow 0$ and slowly decreases with $x$ for $x>1$ (for $x = 0.2$ and 10,  $F(x) = 0.5$ and 0.6, respectively) \cite{Micklitz:2006_PRL}.  The $r$ value is not universal, but varies with the hole position in respect to the edge and QW center. To estimate $L_{\text{p}}$ we adopt \cite{Dietl:2023_PRB}: $N_h =0.5\cdot10^{11}$\,cm$^{-2}$, an average value of $r^{(i)}$ as $r_{Dx} = 0.13$ , the cut-off length beyond which strong coupling of holes and electrons tends to vanish $y_{\text{c}} = 2b = 10$\,nm, and an average value of $f(T/T_{\text{K}}) = 0.4$. These numbers lead to  $L_{\text{p}} = 4$\,$\mu$m, the order of magnitude consistent with experimental findings \cite{Konig:2007_S,Lunczer:2019_PRL,Majewicz:2019_PhD}. A more elaborated approach \cite{Dietl:2023_PRB} provides conductance values and temperature dependence $G(T)$ that agree with experimental observations, if  the influence of Luttinger correlation effects upon $r$ \cite{Vayrynen:2016_PRB} is taken into account.

Equation \ref{eq:Lp} implies that $L_{\text{p}}^{-1}$ scales linearly with $N_h$. This fact explains a two orders of magnitude longer $L_{\text{p}}$ in  HgTe QWs \cite{Konig:2007_S,Lunczer:2019_PRL,Majewicz:2019_PhD} compared to 1T'-WTe$_2$ 2D monolayers \cite{Fei:2017_NP,Wu:2018_S}, as gating experiments point to correspondingly different acceptor concentrations in these two systems $N_{\text{a}} =10^{11}$ and $10^{13}$\,cm$^{-2}$, respectively \cite{Bendias:2018_NL,Yahniuk:2021_arXiv,Fei:2017_NP}. Furthermore, a small number of relevant acceptor holes leads to reproducible resistance fluctuations \cite{Konig:2007_S,Shamim:2021_NC,Fei:2017_NP,Wu:2018_S}. At the same time, a decrease of conductance seen in scanning gate microscopy experiments \cite{Konig:2013_PRX} results from a local increase in the number of acceptor holes in the edge region. Filamentary charging and discharging of barrier acceptors under a strong gate electric field may account for hystereses and irreversibilities in low-temperature transport properties when cycling the gate voltage \cite{Lunczer:2019_PRL,Shamim:2021_NC}.

An unexpected appearance of quantized resistance below 0.3\,K in a Hg$_{0.988}$Mn$_{0.012}$Te QW \cite{Shamim:2021_NC} can be elucidated using the acceptor model by spin splitting $\Delta$ of hole states originating from the BMP effect, as for $x = 0.012$, $\Delta > k_{\text{B}}T$ at $T < 3.5$\,K \cite{Dietl:2023_PRB}. Interestingly, the existing theories on the disappearance of the Kondo effect in a magnetic field assume the same $\Delta$ for the impurity and band states \cite{Filippone:2018_PRB}, which is not the case in the presence of BMPs.

{\em{Conclusions and outlook}}--The proposed impurity band model can elucidate the critical properties of QSHE materials. In addition to controlling carriers' densities and mobilities, the charge dopants enlarge the spin Hall plateau width, but diminish the quantization precision. However, the resistance quantization accuracy can be recovered by doping topological QWs with isoelectronic magnetic impurities, as the formation of the bound magnetic polarons weakens the Kondo effect. Similarly, impurities with a negative value of Hubbard's $U$ can pin the Fermi level in the gap, but will not contribute to backscattering, provided two trapped carriers form a spin singlet. Even if such DX$^-$  or AX$^+$ centers are unstable under ambient conditions \cite{Chatratin:2023_JPCL}, fast gate sweeping, light, or hydrostatic pressure might serve for their activation \cite{Suski:1990_SST}.

In this Letter, the model's quantitative predictions have been compared to experimental data on HgTe and Hg$_{1-x}$Mn$_{x}$Te QWs as well as on 1T'-WTe$_2$ 2D monolayers, however, it would be interesting to verify the model in the case of other QSHE candidate materials, such as $\alpha$-Sn and Bi films, other 2D monolayers, and Heusler compounds with an inverted band structure. More generally, while electrostatic gating is widely used to reveal the unique properties of quantum materials, the results presented here demonstrate that charge dopants play an important and unanticipated role in the physics and applications of topological semiconductors. Finally, we mention that our theory have been limited to the Ohmic range. A pallet of new phenomena is expected beyond the linear response regime \cite{Vayrynen:2016_PRB,Lunde:2012_PRB,Del_Maestro:2013_PRB}.

{\em{Acknowledgments}}--This project was supported  by the Foundation for Polish Science through the International Research
Agendas program co-financed by the European Union within the Smart Growth Operational Programme (MAB/2017/1).


\begin{thebibliography}{49}%
\makeatletter
\providecommand \@ifxundefined [1]{%
 \@ifx{#1\undefined}
}%
\providecommand \@ifnum [1]{%
 \ifnum #1\expandafter \@firstoftwo
 \else \expandafter \@secondoftwo
 \fi
}%
\providecommand \@ifx [1]{%
 \ifx #1\expandafter \@firstoftwo
 \else \expandafter \@secondoftwo
 \fi
}%
\providecommand \natexlab [1]{#1}%
\providecommand \enquote  [1]{``#1''}%
\providecommand \bibnamefont  [1]{#1}%
\providecommand \bibfnamefont [1]{#1}%
\providecommand \citenamefont [1]{#1}%
\providecommand \href@noop [0]{\@secondoftwo}%
\providecommand \href [0]{\begingroup \@sanitize@url \@href}%
\providecommand \@href[1]{\@@startlink{#1}\@@href}%
\providecommand \@@href[1]{\endgroup#1\@@endlink}%
\providecommand \@sanitize@url [0]{\catcode `\\12\catcode `\$12\catcode
  `\&12\catcode `\#12\catcode `\^12\catcode `\_12\catcode `\%12\relax}%
\providecommand \@@startlink[1]{}%
\providecommand \@@endlink[0]{}%
\providecommand \url  [0]{\begingroup\@sanitize@url \@url }%
\providecommand \@url [1]{\endgroup\@href {#1}{\urlprefix }}%
\providecommand \urlprefix  [0]{URL }%
\providecommand \Eprint [0]{\href }%
\providecommand \doibase [0]{http://dx.doi.org/}%
\providecommand \selectlanguage [0]{\@gobble}%
\providecommand \bibinfo  [0]{\@secondoftwo}%
\providecommand \bibfield  [0]{\@secondoftwo}%
\providecommand \translation [1]{[#1]}%
\providecommand \BibitemOpen [0]{}%
\providecommand \bibitemStop [0]{}%
\providecommand \bibitemNoStop [0]{.\EOS\space}%
\providecommand \EOS [0]{\spacefactor3000\relax}%
\providecommand \BibitemShut  [1]{\csname bibitem#1\endcsname}%
\let\auto@bib@innerbib\@empty
\bibitem [{\citenamefont {{von Klitzing}}\ \emph {et~al.}(2020)\citenamefont
  {{von Klitzing}}, \citenamefont {Chakraborty}, \citenamefont {Kim},
  \citenamefont {Madhavan}, \citenamefont {Dai}, \citenamefont {McIver},
  \citenamefont {Tokura}, \citenamefont {Savary}, \citenamefont {Smirnova},
  \citenamefont {Rey}, \citenamefont {Felser}, \citenamefont {Gooth},\ and\
  \citenamefont {Qi}}]{Klitzing:2020_NRP}%
  \BibitemOpen
  \bibfield  {author} {\bibinfo {author} {\bibfnamefont {K.}~\bibnamefont {{von
  Klitzing}}}, \bibinfo {author} {\bibfnamefont {T.}~\bibnamefont
  {Chakraborty}}, \bibinfo {author} {\bibfnamefont {P.}~\bibnamefont {Kim}},
  \bibinfo {author} {\bibfnamefont {V.}~\bibnamefont {Madhavan}}, \bibinfo
  {author} {\bibfnamefont {Xi}~\bibnamefont {Dai}}, \bibinfo {author}
  {\bibfnamefont {J.}~\bibnamefont {McIver}}, \bibinfo {author} {\bibfnamefont
  {Y.}~\bibnamefont {Tokura}}, \bibinfo {author} {\bibfnamefont
  {L.}~\bibnamefont {Savary}}, \bibinfo {author} {\bibfnamefont
  {D.}~\bibnamefont {Smirnova}}, \bibinfo {author} {\bibfnamefont {A.~M.}\
  \bibnamefont {Rey}}, \bibinfo {author} {\bibfnamefont {C.}~\bibnamefont
  {Felser}}, \bibinfo {author} {\bibfnamefont {J.}~\bibnamefont {Gooth}}, \
  and\ \bibinfo {author} {\bibfnamefont {Xiaoliang}\ \bibnamefont {Qi}},\
  }\bibfield  {title} {\enquote {\bibinfo {title} {40 years of the quantum
  {Hall} effect},}\ }\href {\doibase 10.1038/s42254-020-0209-1} {\bibfield
  {journal} {\bibinfo  {journal} {Nat. Rev. Phys.}\ }\textbf {\bibinfo {volume}
  {2}},\ \bibinfo {pages} {397--401} (\bibinfo {year} {2020})}\BibitemShut
  {NoStop}%
\bibitem [{\citenamefont {Rigosi}\ and\ \citenamefont
  {Elmquist}(2019)}]{Rigosi:2019_SST}%
  \BibitemOpen
  \bibfield  {author} {\bibinfo {author} {\bibfnamefont {A.~F.}\ \bibnamefont
  {Rigosi}}\ and\ \bibinfo {author} {\bibfnamefont {R.~E.}\ \bibnamefont
  {Elmquist}},\ }\bibfield  {title} {\enquote {\bibinfo {title} {The quantum
  {Hall} effect in the era of the new {SI}},}\ }\href {\doibase
  10.1088/1361-6641/ab37d3} {\bibfield  {journal} {\bibinfo  {journal}
  {Semicon. Sci. Techn.}\ }\textbf {\bibinfo {volume} {34}},\ \bibinfo {pages}
  {093004} (\bibinfo {year} {2019})}\BibitemShut {NoStop}%
\bibitem [{\citenamefont {Kane}\ and\ \citenamefont
  {Mele}(2005)}]{Kane:2005_PRL}%
  \BibitemOpen
  \bibfield  {author} {\bibinfo {author} {\bibfnamefont {C.~L.}\ \bibnamefont
  {Kane}}\ and\ \bibinfo {author} {\bibfnamefont {E.~J.}\ \bibnamefont
  {Mele}},\ }\bibfield  {title} {\enquote {\bibinfo {title} {Quantum spin
  {Hall} effect in graphene},}\ }\href {\doibase 10.1103/PhysRevLett.95.226801}
  {\bibfield  {journal} {\bibinfo  {journal} {Phys. Rev. Lett.}\ }\textbf
  {\bibinfo {volume} {95}},\ \bibinfo {pages} {226801} (\bibinfo {year}
  {2005})}\BibitemShut {NoStop}%
\bibitem [{\citenamefont {Bernevig}\ \emph {et~al.}(2006)\citenamefont
  {Bernevig}, \citenamefont {Hughes},\ and\ \citenamefont
  {Zhang}}]{Bernevig:2006_S}%
  \BibitemOpen
  \bibfield  {author} {\bibinfo {author} {\bibfnamefont {B.~A.}\ \bibnamefont
  {Bernevig}}, \bibinfo {author} {\bibfnamefont {T.~L.}\ \bibnamefont
  {Hughes}}, \ and\ \bibinfo {author} {\bibfnamefont {Shou-Cheng}\ \bibnamefont
  {Zhang}},\ }\bibfield  {title} {\enquote {\bibinfo {title} {Quantum spin
  {Hall} effect and topological phase transition in {HgTe} quantum wells},}\
  }\href {\doibase 10.1126/science.1133734} {\bibfield  {journal} {\bibinfo
  {journal} {Science}\ }\textbf {\bibinfo {volume} {314}},\ \bibinfo {pages}
  {1757--1761} (\bibinfo {year} {2006})}\BibitemShut {NoStop}%
\bibitem [{\citenamefont {K\"{o}nig}\ \emph {et~al.}(2007)\citenamefont
  {K\"{o}nig}, \citenamefont {Wiedmann}, \citenamefont {Br\"{u}ne},
  \citenamefont {Roth}, \citenamefont {Buhmann}, \citenamefont {Molenkamp},
  \citenamefont {Qi},\ and\ \citenamefont {Zhang}}]{Konig:2007_S}%
  \BibitemOpen
  \bibfield  {author} {\bibinfo {author} {\bibfnamefont {M.}~\bibnamefont
  {K\"{o}nig}}, \bibinfo {author} {\bibfnamefont {S.}~\bibnamefont {Wiedmann}},
  \bibinfo {author} {\bibfnamefont {C.}~\bibnamefont {Br\"{u}ne}}, \bibinfo
  {author} {\bibfnamefont {A.}~\bibnamefont {Roth}}, \bibinfo {author}
  {\bibfnamefont {H.}~\bibnamefont {Buhmann}}, \bibinfo {author} {\bibfnamefont
  {L.~W.}\ \bibnamefont {Molenkamp}}, \bibinfo {author} {\bibfnamefont
  {Xiao-Liang}\ \bibnamefont {Qi}}, \ and\ \bibinfo {author} {\bibfnamefont
  {Shou-Cheng}\ \bibnamefont {Zhang}},\ }\bibfield  {title} {\enquote {\bibinfo
  {title} {Quantum spin {Hall} insulator state in {HgTe} quantum wells},}\
  }\href {\doibase 10.1126/science.1148047} {\bibfield  {journal} {\bibinfo
  {journal} {Science}\ }\textbf {\bibinfo {volume} {318}},\ \bibinfo {pages}
  {766--770} (\bibinfo {year} {2007})}\BibitemShut {NoStop}%
\bibitem [{\citenamefont {Bendias}\ \emph {et~al.}(2018)\citenamefont
  {Bendias}, \citenamefont {Shamim}, \citenamefont {Herrmann}, \citenamefont
  {Budewitz}, \citenamefont {Shekhar}, \citenamefont {Leubner}, \citenamefont
  {Kleinlein}, \citenamefont {Bocquillon}, \citenamefont {Buhmann},\ and\
  \citenamefont {Molenkamp}}]{Bendias:2018_NL}%
  \BibitemOpen
  \bibfield  {author} {\bibinfo {author} {\bibfnamefont {K.}~\bibnamefont
  {Bendias}}, \bibinfo {author} {\bibfnamefont {S.}~\bibnamefont {Shamim}},
  \bibinfo {author} {\bibfnamefont {O.}~\bibnamefont {Herrmann}}, \bibinfo
  {author} {\bibfnamefont {A.}~\bibnamefont {Budewitz}}, \bibinfo {author}
  {\bibfnamefont {P.}~\bibnamefont {Shekhar}}, \bibinfo {author} {\bibfnamefont
  {P.}~\bibnamefont {Leubner}}, \bibinfo {author} {\bibfnamefont
  {J.}~\bibnamefont {Kleinlein}}, \bibinfo {author} {\bibfnamefont
  {E.}~\bibnamefont {Bocquillon}}, \bibinfo {author} {\bibfnamefont
  {H.}~\bibnamefont {Buhmann}}, \ and\ \bibinfo {author} {\bibfnamefont
  {L.~W.}\ \bibnamefont {Molenkamp}},\ }\bibfield  {title} {\enquote {\bibinfo
  {title} {High mobility {HgTe} microstructures for quantum spin {Hall}
  studies},}\ }\href {\doibase 10.1021/acs.nanolett.8b01405} {\bibfield
  {journal} {\bibinfo  {journal} {Nano Lett.}\ }\textbf {\bibinfo {volume}
  {18}},\ \bibinfo {pages} {4831--4836} (\bibinfo {year} {2018})}\BibitemShut
  {NoStop}%
\bibitem [{\citenamefont {Shamim}\ \emph {et~al.}(2021)\citenamefont {Shamim},
  \citenamefont {Beugeling}, \citenamefont {Shekhar}, \citenamefont {Bendias},
  \citenamefont {Lunczer}, \citenamefont {Kleinlein}, \citenamefont {Buhmann},\
  and\ \citenamefont {Molenkamp}}]{Shamim:2021_NC}%
  \BibitemOpen
  \bibfield  {author} {\bibinfo {author} {\bibfnamefont {S.}~\bibnamefont
  {Shamim}}, \bibinfo {author} {\bibfnamefont {W.}~\bibnamefont {Beugeling}},
  \bibinfo {author} {\bibfnamefont {P.}~\bibnamefont {Shekhar}}, \bibinfo
  {author} {\bibfnamefont {K.}~\bibnamefont {Bendias}}, \bibinfo {author}
  {\bibfnamefont {L.}~\bibnamefont {Lunczer}}, \bibinfo {author} {\bibfnamefont
  {J.}~\bibnamefont {Kleinlein}}, \bibinfo {author} {\bibfnamefont
  {H.}~\bibnamefont {Buhmann}}, \ and\ \bibinfo {author} {\bibfnamefont
  {L.~W.}\ \bibnamefont {Molenkamp}},\ }\bibfield  {title} {\enquote {\bibinfo
  {title} {Quantized spin {Hall} conductance in a magnetically doped two
  dimensional topological insulator},}\ }\href {\doibase
  10.1038/s41467-021-23262-1} {\bibfield  {journal} {\bibinfo  {journal} {Nat.
  Commun.}\ }\textbf {\bibinfo {volume} {12}},\ \bibinfo {pages} {3193}
  (\bibinfo {year} {2021})}\BibitemShut {NoStop}%
\bibitem [{\citenamefont {Fei}\ \emph {et~al.}(2017)\citenamefont {Fei},
  \citenamefont {Palomaki}, \citenamefont {Wu}, \citenamefont {Zhao},
  \citenamefont {Cai}, \citenamefont {Sun}, \citenamefont {Nguyen},
  \citenamefont {Finney}, \citenamefont {Xu},\ and\ \citenamefont
  {Cobden}}]{Fei:2017_NP}%
  \BibitemOpen
  \bibfield  {author} {\bibinfo {author} {\bibfnamefont {Zaiyao}\ \bibnamefont
  {Fei}}, \bibinfo {author} {\bibfnamefont {T.}~\bibnamefont {Palomaki}},
  \bibinfo {author} {\bibfnamefont {Sanfeng}\ \bibnamefont {Wu}}, \bibinfo
  {author} {\bibfnamefont {Wenjin}\ \bibnamefont {Zhao}}, \bibinfo {author}
  {\bibfnamefont {Xinghan}\ \bibnamefont {Cai}}, \bibinfo {author}
  {\bibfnamefont {Bosong}\ \bibnamefont {Sun}}, \bibinfo {author}
  {\bibfnamefont {Paul}\ \bibnamefont {Nguyen}}, \bibinfo {author}
  {\bibfnamefont {J.}~\bibnamefont {Finney}}, \bibinfo {author} {\bibfnamefont
  {Xiaodong}\ \bibnamefont {Xu}}, \ and\ \bibinfo {author} {\bibfnamefont
  {D.~H.}\ \bibnamefont {Cobden}},\ }\bibfield  {title} {\enquote {\bibinfo
  {title} {Edge conduction in monolayer {WTe$_2$}},}\ }\href {\doibase
  10.1038/nphys4091} {\bibfield  {journal} {\bibinfo  {journal} {Nat. Phys.}\
  }\textbf {\bibinfo {volume} {13}},\ \bibinfo {pages} {677--682} (\bibinfo
  {year} {2017})}\BibitemShut {NoStop}%
\bibitem [{\citenamefont {Wu}\ \emph {et~al.}(2018)\citenamefont {Wu},
  \citenamefont {Fatemi}, \citenamefont {Gibson}, \citenamefont {Watanabe},
  \citenamefont {Taniguchi}, \citenamefont {Cava},\ and\ \citenamefont
  {Jarillo-Herrero}}]{Wu:2018_S}%
  \BibitemOpen
  \bibfield  {author} {\bibinfo {author} {\bibfnamefont {Sanfeng}\ \bibnamefont
  {Wu}}, \bibinfo {author} {\bibfnamefont {V.}~\bibnamefont {Fatemi}}, \bibinfo
  {author} {\bibfnamefont {Q.~D.}\ \bibnamefont {Gibson}}, \bibinfo {author}
  {\bibfnamefont {K.}~\bibnamefont {Watanabe}}, \bibinfo {author}
  {\bibfnamefont {T.}~\bibnamefont {Taniguchi}}, \bibinfo {author}
  {\bibfnamefont {R.~J.}\ \bibnamefont {Cava}}, \ and\ \bibinfo {author}
  {\bibfnamefont {P.}~\bibnamefont {Jarillo-Herrero}},\ }\bibfield  {title}
  {\enquote {\bibinfo {title} {Observation of the quantum spin {Hall} effect up
  to 100 kelvin in a monolayer crystal},}\ }\href {\doibase
  10.1126/science.aan6003} {\bibfield  {journal} {\bibinfo  {journal}
  {Science}\ }\textbf {\bibinfo {volume} {359}},\ \bibinfo {pages} {76--79}
  (\bibinfo {year} {2018})}\BibitemShut {NoStop}%
\bibitem [{\citenamefont {Hsu}\ \emph {et~al.}(2021)\citenamefont {Hsu},
  \citenamefont {Stano}, \citenamefont {Klinovaja},\ and\ \citenamefont
  {Loss}}]{Hsu:2021_SST}%
  \BibitemOpen
  \bibfield  {author} {\bibinfo {author} {\bibfnamefont {Chen-Hsuan}\
  \bibnamefont {Hsu}}, \bibinfo {author} {\bibfnamefont {P.}~\bibnamefont
  {Stano}}, \bibinfo {author} {\bibfnamefont {J.}~\bibnamefont {Klinovaja}}, \
  and\ \bibinfo {author} {\bibfnamefont {D.}~\bibnamefont {Loss}},\ }\bibfield
  {title} {\enquote {\bibinfo {title} {Helical liquids in semiconductors},}\
  }\href {\doibase 10.1088/1361-6641/ac2c27} {\bibfield  {journal} {\bibinfo
  {journal} {Semicon. Sci. Technol.}\ }\textbf {\bibinfo {volume} {36}},\
  \bibinfo {pages} {123003} (\bibinfo {year} {2021})}\BibitemShut {NoStop}%
\bibitem [{\citenamefont {Yevtushenko}\ and\ \citenamefont
  {Yudson}(2022)}]{Yevtushenko:2022_NJP}%
  \BibitemOpen
  \bibfield  {author} {\bibinfo {author} {\bibfnamefont {O.~M.}\ \bibnamefont
  {Yevtushenko}}\ and\ \bibinfo {author} {\bibfnamefont {V.~I.}\ \bibnamefont
  {Yudson}},\ }\bibfield  {title} {\enquote {\bibinfo {title} {Protection of
  edge transport in quantum spin {Hall} samples: spin-symmetry based general
  approach and examples},}\ }\href {\doibase 10.1088/1367-2630/ac50e9}
  {\bibfield  {journal} {\bibinfo  {journal} {New J. Phys.}\ }\textbf {\bibinfo
  {volume} {24}},\ \bibinfo {pages} {023040} (\bibinfo {year}
  {2022})}\BibitemShut {NoStop}%
\bibitem [{\citenamefont {V\"ayrynen}\ \emph {et~al.}(2014)\citenamefont
  {V\"ayrynen}, \citenamefont {Goldstein}, \citenamefont {Gefen},\ and\
  \citenamefont {Glazman}}]{Vayrynen:2014_PRB}%
  \BibitemOpen
  \bibfield  {author} {\bibinfo {author} {\bibfnamefont {J.~I.}\ \bibnamefont
  {V\"ayrynen}}, \bibinfo {author} {\bibfnamefont {M.}~\bibnamefont
  {Goldstein}}, \bibinfo {author} {\bibfnamefont {Y.}~\bibnamefont {Gefen}}, \
  and\ \bibinfo {author} {\bibfnamefont {L.~I.}\ \bibnamefont {Glazman}},\
  }\bibfield  {title} {\enquote {\bibinfo {title} {Resistance of helical edges
  formed in a semiconductor heterostructure},}\ }\href {\doibase
  10.1103/PhysRevB.90.115309} {\bibfield  {journal} {\bibinfo  {journal} {Phys.
  Rev. B}\ }\textbf {\bibinfo {volume} {90}},\ \bibinfo {pages} {115309}
  (\bibinfo {year} {2014})}\BibitemShut {NoStop}%
\bibitem [{\citenamefont {Grabecki}\ \emph {et~al.}(2013)\citenamefont
  {Grabecki}, \citenamefont {Wr{\'o}bel}, \citenamefont {Czapkiewicz},
  \citenamefont {Cywi\'nski}, \citenamefont {Giera{\l}towska}, \citenamefont
  {Guziewicz}, \citenamefont {Zholudev}, \citenamefont {Gavrilenko},
  \citenamefont {Mikhailov}, \citenamefont {Dvoretski}, \citenamefont {Teppe},
  \citenamefont {Knap},\ and\ \citenamefont {Dietl}}]{Grabecki:2013_PRB}%
  \BibitemOpen
  \bibfield  {author} {\bibinfo {author} {\bibfnamefont {G.}~\bibnamefont
  {Grabecki}}, \bibinfo {author} {\bibfnamefont {J.}~\bibnamefont
  {Wr{\'o}bel}}, \bibinfo {author} {\bibfnamefont {M.}~\bibnamefont
  {Czapkiewicz}}, \bibinfo {author} {\bibfnamefont {{\L}.}~\bibnamefont
  {Cywi\'nski}}, \bibinfo {author} {\bibfnamefont {S.}~\bibnamefont
  {Giera{\l}towska}}, \bibinfo {author} {\bibfnamefont {E.}~\bibnamefont
  {Guziewicz}}, \bibinfo {author} {\bibfnamefont {M.}~\bibnamefont {Zholudev}},
  \bibinfo {author} {\bibfnamefont {V.}~\bibnamefont {Gavrilenko}}, \bibinfo
  {author} {\bibfnamefont {N.~N.}\ \bibnamefont {Mikhailov}}, \bibinfo {author}
  {\bibfnamefont {S.~A.}\ \bibnamefont {Dvoretski}}, \bibinfo {author}
  {\bibfnamefont {F.}~\bibnamefont {Teppe}}, \bibinfo {author} {\bibfnamefont
  {W.}~\bibnamefont {Knap}}, \ and\ \bibinfo {author} {\bibfnamefont
  {T.}~\bibnamefont {Dietl}},\ }\bibfield  {title} {\enquote {\bibinfo {title}
  {Nonlocal resistance and its fluctuations in microstructures of band-inverted
  {HgTe/(Hg,Cd)Te} quantum wells},}\ }\href {\doibase
  10.1103/PhysRevB.88.165309} {\bibfield  {journal} {\bibinfo  {journal} {Phys.
  Rev. B}\ }\textbf {\bibinfo {volume} {88}},\ \bibinfo {pages} {165309}
  (\bibinfo {year} {2013})}\BibitemShut {NoStop}%
\bibitem [{\citenamefont {Lunczer}\ \emph {et~al.}(2019)\citenamefont
  {Lunczer}, \citenamefont {Leubner}, \citenamefont {Endres}, \citenamefont
  {M\"uller}, \citenamefont {Br\"une}, \citenamefont {Buhmann},\ and\
  \citenamefont {Molenkamp}}]{Lunczer:2019_PRL}%
  \BibitemOpen
  \bibfield  {author} {\bibinfo {author} {\bibfnamefont {L.}~\bibnamefont
  {Lunczer}}, \bibinfo {author} {\bibfnamefont {P.}~\bibnamefont {Leubner}},
  \bibinfo {author} {\bibfnamefont {M.}~\bibnamefont {Endres}}, \bibinfo
  {author} {\bibfnamefont {V.~L.}\ \bibnamefont {M\"uller}}, \bibinfo {author}
  {\bibfnamefont {C.}~\bibnamefont {Br\"une}}, \bibinfo {author} {\bibfnamefont
  {H.}~\bibnamefont {Buhmann}}, \ and\ \bibinfo {author} {\bibfnamefont
  {L.~W.}\ \bibnamefont {Molenkamp}},\ }\bibfield  {title} {\enquote {\bibinfo
  {title} {Approaching quantization in macroscopic quantum spin {Hall} devices
  through gate training},}\ }\href {\doibase 10.1103/PhysRevLett.123.047701}
  {\bibfield  {journal} {\bibinfo  {journal} {Phys. Rev. Lett.}\ }\textbf
  {\bibinfo {volume} {123}},\ \bibinfo {pages} {047701} (\bibinfo {year}
  {2019})}\BibitemShut {NoStop}%
\bibitem [{\citenamefont {Wilamowski}\ \emph {et~al.}(1990)\citenamefont
  {Wilamowski}, \citenamefont {{\'S}wi{\c{a}}tek}, \citenamefont {Dietl},\ and\
  \citenamefont {Kossut}}]{Wilamowski:1990_SSC}%
  \BibitemOpen
  \bibfield  {author} {\bibinfo {author} {\bibfnamefont {Z.}~\bibnamefont
  {Wilamowski}}, \bibinfo {author} {\bibfnamefont {K.}~\bibnamefont
  {{\'S}wi{\c{a}}tek}}, \bibinfo {author} {\bibfnamefont {T.}~\bibnamefont
  {Dietl}}, \ and\ \bibinfo {author} {\bibfnamefont {J.}~\bibnamefont
  {Kossut}},\ }\bibfield  {title} {\enquote {\bibinfo {title} {Resonant states
  in semiconductors: A quantitative study of {HgSe:Fe}},}\ }\href {\doibase
  https://doi.org/10.1016/0038-1098(90)90945-8} {\bibfield  {journal} {\bibinfo
   {journal} {Solid State Commun.}\ }\textbf {\bibinfo {volume} {74}},\
  \bibinfo {pages} {833--837} (\bibinfo {year} {1990})}\BibitemShut {NoStop}%
\bibitem [{\citenamefont {Shamim}\ \emph {et~al.}(2020)\citenamefont {Shamim},
  \citenamefont {Beugeling}, \citenamefont {B\"ottcher}, \citenamefont
  {Shekhar}, \citenamefont {Budewitz}, \citenamefont {Leubner}, \citenamefont
  {Lunczer}, \citenamefont {Hankiewicz}, \citenamefont {Buhmann},\ and\
  \citenamefont {Molenkamp}}]{Shamim:2020_SA}%
  \BibitemOpen
  \bibfield  {author} {\bibinfo {author} {\bibfnamefont {S.}~\bibnamefont
  {Shamim}}, \bibinfo {author} {\bibfnamefont {W.}~\bibnamefont {Beugeling}},
  \bibinfo {author} {\bibfnamefont {J.}~\bibnamefont {B\"ottcher}}, \bibinfo
  {author} {\bibfnamefont {P.}~\bibnamefont {Shekhar}}, \bibinfo {author}
  {\bibfnamefont {A.}~\bibnamefont {Budewitz}}, \bibinfo {author}
  {\bibfnamefont {P.}~\bibnamefont {Leubner}}, \bibinfo {author} {\bibfnamefont
  {L.}~\bibnamefont {Lunczer}}, \bibinfo {author} {\bibfnamefont {E.~M.}\
  \bibnamefont {Hankiewicz}}, \bibinfo {author} {\bibfnamefont
  {H.}~\bibnamefont {Buhmann}}, \ and\ \bibinfo {author} {\bibfnamefont
  {L.~W.}\ \bibnamefont {Molenkamp}},\ }\bibfield  {title} {\enquote {\bibinfo
  {title} {Emergent quantum {Hall} effects below 50 {mT} in a two-dimensional
  topological insulator},}\ }\href {\doibase 10.1126/sciadv.aba4625} {\bibfield
   {journal} {\bibinfo  {journal} {Adv. Sci.}\ }\textbf {\bibinfo {volume}
  {6}},\ \bibinfo {pages} {eaba4625} (\bibinfo {year} {2020})}\BibitemShut
  {NoStop}%
\bibitem [{\citenamefont {Yahniuk}\ \emph {et~al.}()\citenamefont {Yahniuk},
  \citenamefont {Kazakov}, \citenamefont {Jouault}, \citenamefont
  {Krishtopenko}, \citenamefont {Kret}, \citenamefont {Grabecki}, \citenamefont
  {Cywi\'nski}, \citenamefont {Mikhailov}, \citenamefont {Dvoretskii},
  \citenamefont {Przybytek}, \citenamefont {Gavrilenko}, \citenamefont {Teppe},
  \citenamefont {Dietl},\ and\ \citenamefont {Knap}}]{Yahniuk:2021_arXiv}%
  \BibitemOpen
  \bibfield  {author} {\bibinfo {author} {\bibfnamefont {I.}~\bibnamefont
  {Yahniuk}}, \bibinfo {author} {\bibfnamefont {A.}~\bibnamefont {Kazakov}},
  \bibinfo {author} {\bibfnamefont {B.}~\bibnamefont {Jouault}}, \bibinfo
  {author} {\bibfnamefont {S.~S.}\ \bibnamefont {Krishtopenko}}, \bibinfo
  {author} {\bibfnamefont {S.}~\bibnamefont {Kret}}, \bibinfo {author}
  {\bibfnamefont {G.}~\bibnamefont {Grabecki}}, \bibinfo {author}
  {\bibfnamefont {G.}~\bibnamefont {Cywi\'nski}}, \bibinfo {author}
  {\bibfnamefont {N.~N.}\ \bibnamefont {Mikhailov}}, \bibinfo {author}
  {\bibfnamefont {S.~A.}\ \bibnamefont {Dvoretskii}}, \bibinfo {author}
  {\bibfnamefont {J.}~\bibnamefont {Przybytek}}, \bibinfo {author}
  {\bibfnamefont {V.~I.}\ \bibnamefont {Gavrilenko}}, \bibinfo {author}
  {\bibfnamefont {F.}~\bibnamefont {Teppe}}, \bibinfo {author} {\bibfnamefont
  {T.}~\bibnamefont {Dietl}}, \ and\ \bibinfo {author} {\bibfnamefont
  {W.}~\bibnamefont {Knap}},\ }\bibfield  {title} {\enquote {\bibinfo {title}
  {{HgTe} quantum wells for {QHE} metrology under soft cryomagnetic conditions:
  permanent magnets and liquid {$^4$He} temperatures},}\ }\href {\doibase
  10.48550/arXiv.2111.07581} {\ 10.48550/arXiv.2111.07581}\BibitemShut
  {NoStop}%
\bibitem [{\citenamefont {Dartiailh}\ \emph {et~al.}(2020)\citenamefont
  {Dartiailh}, \citenamefont {Hartinger}, \citenamefont {Gourmelon},
  \citenamefont {Bendias}, \citenamefont {Bartolomei}, \citenamefont {Kamata},
  \citenamefont {Berroir}, \citenamefont {F\`eve}, \citenamefont
  {Pla{\c{c}}ais}, \citenamefont {Lunczer}, \citenamefont {Schlereth},
  \citenamefont {Buhmann}, \citenamefont {Molenkamp},\ and\ \citenamefont
  {Bocquillon}}]{Dartiailh:2020_PRL}%
  \BibitemOpen
  \bibfield  {author} {\bibinfo {author} {\bibfnamefont {M.~C.}\ \bibnamefont
  {Dartiailh}}, \bibinfo {author} {\bibfnamefont {S.}~\bibnamefont
  {Hartinger}}, \bibinfo {author} {\bibfnamefont {A.}~\bibnamefont
  {Gourmelon}}, \bibinfo {author} {\bibfnamefont {K.}~\bibnamefont {Bendias}},
  \bibinfo {author} {\bibfnamefont {H.}~\bibnamefont {Bartolomei}}, \bibinfo
  {author} {\bibfnamefont {H.}~\bibnamefont {Kamata}}, \bibinfo {author}
  {\bibfnamefont {J.-M.}\ \bibnamefont {Berroir}}, \bibinfo {author}
  {\bibfnamefont {G.}~\bibnamefont {F\`eve}}, \bibinfo {author} {\bibfnamefont
  {B.}~\bibnamefont {Pla{\c{c}}ais}}, \bibinfo {author} {\bibfnamefont
  {L.}~\bibnamefont {Lunczer}}, \bibinfo {author} {\bibfnamefont
  {R.}~\bibnamefont {Schlereth}}, \bibinfo {author} {\bibfnamefont
  {H.}~\bibnamefont {Buhmann}}, \bibinfo {author} {\bibfnamefont {L.~W.}\
  \bibnamefont {Molenkamp}}, \ and\ \bibinfo {author} {\bibfnamefont
  {E.}~\bibnamefont {Bocquillon}},\ }\bibfield  {title} {\enquote {\bibinfo
  {title} {Dynamical separation of bulk and edge transport in {HgTe}-based {2D}
  topological insulators},}\ }\href {\doibase 10.1103/PhysRevLett.124.076802}
  {\bibfield  {journal} {\bibinfo  {journal} {Phys. Rev. Lett.}\ }\textbf
  {\bibinfo {volume} {124}},\ \bibinfo {pages} {076802} (\bibinfo {year}
  {2020})}\BibitemShut {NoStop}%
\bibitem [{\citenamefont {\'{S}liwa}\ and\ \citenamefont
  {Dietl}(2008)}]{Sliwa:2008_PRB}%
  \BibitemOpen
  \bibfield  {author} {\bibinfo {author} {\bibfnamefont {C.}~\bibnamefont
  {\'{S}liwa}}\ and\ \bibinfo {author} {\bibfnamefont {T.}~\bibnamefont
  {Dietl}},\ }\bibfield  {title} {\enquote {\bibinfo {title} {Electron-hole
  contribution to the apparent $s\ensuremath{-}d$ exchange interaction in
  {III-V} dilute magnetic semiconductors},}\ }\href {\doibase
  10.1103/PhysRevB.78.165205} {\bibfield  {journal} {\bibinfo  {journal} {Phys.
  Rev. B}\ }\textbf {\bibinfo {volume} {78}},\ \bibinfo {pages} {165205}
  (\bibinfo {year} {2008})}\BibitemShut {NoStop}%
\bibitem [{\citenamefont {Maciejko}\ \emph {et~al.}(2009)\citenamefont
  {Maciejko}, \citenamefont {Liu}, \citenamefont {Oreg}, \citenamefont {Qi},
  \citenamefont {Wu},\ and\ \citenamefont {Zhang}}]{Maciejko:2009_PRL}%
  \BibitemOpen
  \bibfield  {author} {\bibinfo {author} {\bibfnamefont {J.}~\bibnamefont
  {Maciejko}}, \bibinfo {author} {\bibfnamefont {Chaoxing}\ \bibnamefont
  {Liu}}, \bibinfo {author} {\bibfnamefont {Y.}~\bibnamefont {Oreg}}, \bibinfo
  {author} {\bibfnamefont {Xiao-Liang}\ \bibnamefont {Qi}}, \bibinfo {author}
  {\bibfnamefont {Congjun}\ \bibnamefont {Wu}}, \ and\ \bibinfo {author}
  {\bibfnamefont {Shou-Cheng}\ \bibnamefont {Zhang}},\ }\bibfield  {title}
  {\enquote {\bibinfo {title} {Kondo effect in the helical edge liquid of the
  quantum spin {Hall} state},}\ }\href {\doibase
  10.1103/PhysRevLett.102.256803} {\bibfield  {journal} {\bibinfo  {journal}
  {Phys. Rev. Lett.}\ }\textbf {\bibinfo {volume} {102}},\ \bibinfo {pages}
  {256803} (\bibinfo {year} {2009})}\BibitemShut {NoStop}%
\bibitem [{\citenamefont {Tanaka}\ \emph {et~al.}(2011)\citenamefont {Tanaka},
  \citenamefont {Furusaki},\ and\ \citenamefont {Matveev}}]{Tanaka:2011_PRL}%
  \BibitemOpen
  \bibfield  {author} {\bibinfo {author} {\bibfnamefont {Y.}~\bibnamefont
  {Tanaka}}, \bibinfo {author} {\bibfnamefont {A.}~\bibnamefont {Furusaki}}, \
  and\ \bibinfo {author} {\bibfnamefont {K.~A.}\ \bibnamefont {Matveev}},\
  }\bibfield  {title} {\enquote {\bibinfo {title} {Conductance of a helical
  edge liquid coupled to a magnetic impurity},}\ }\href {\doibase
  10.1103/PhysRevLett.106.236402} {\bibfield  {journal} {\bibinfo  {journal}
  {Phys. Rev. Lett.}\ }\textbf {\bibinfo {volume} {106}},\ \bibinfo {pages}
  {236402} (\bibinfo {year} {2011})}\BibitemShut {NoStop}%
\bibitem [{\citenamefont {Micklitz}\ \emph {et~al.}(2006)\citenamefont
  {Micklitz}, \citenamefont {Altland}, \citenamefont {Costi},\ and\
  \citenamefont {Rosch}}]{Micklitz:2006_PRL}%
  \BibitemOpen
  \bibfield  {author} {\bibinfo {author} {\bibfnamefont {T.}~\bibnamefont
  {Micklitz}}, \bibinfo {author} {\bibfnamefont {A.}~\bibnamefont {Altland}},
  \bibinfo {author} {\bibfnamefont {T.~A.}\ \bibnamefont {Costi}}, \ and\
  \bibinfo {author} {\bibfnamefont {A.}~\bibnamefont {Rosch}},\ }\bibfield
  {title} {\enquote {\bibinfo {title} {Universal dephasing rate due to diluted
  {Kondo} impurities},}\ }\href {\doibase 10.1103/PhysRevLett.96.226601}
  {\bibfield  {journal} {\bibinfo  {journal} {Phys. Rev. Lett.}\ }\textbf
  {\bibinfo {volume} {96}},\ \bibinfo {pages} {226601} (\bibinfo {year}
  {2006})}\BibitemShut {NoStop}%
\bibitem [{\citenamefont {Dietl}(2023)}]{Dietl:2023_PRB}%
  \BibitemOpen
  \bibfield  {author} {\bibinfo {author} {\bibfnamefont {T.}~\bibnamefont
  {Dietl}},\ }\bibfield  {title} {\enquote {\bibinfo {title} {Quantitative
  theory of backscattering in topological {HgTe} and {(Hg,Mn)Te} quantum wells:
  acceptor states, {Kondo} effect, precessional dephasing, and bound magnetic
  polaron},}\ }\href {\doibase 10.1103/PhysRevB.107.085421} 
  {\bibfield {journal} {\bibinfo {journal} {Phys. Rev. B}\ }\textbf 
  {\bibinfo {volume} {107}},\ \bibinfo {pages} {085421} (\bibinfo {year} 
  {2023})}\BibitemShut {NoStop}%
\bibitem [{\citenamefont {Umansky}\ \emph {et~al.}(2009)\citenamefont
  {Umansky}, \citenamefont {Heiblum}, \citenamefont {Levinson}, \citenamefont
  {Smet}, \citenamefont {N\"ubler},\ and\ \citenamefont
  {Dolev}}]{Umansky:2009_JCG}%
  \BibitemOpen
  \bibfield  {author} {\bibinfo {author} {\bibfnamefont {V.}~\bibnamefont
  {Umansky}}, \bibinfo {author} {\bibfnamefont {M.}~\bibnamefont {Heiblum}},
  \bibinfo {author} {\bibfnamefont {Y.}~\bibnamefont {Levinson}}, \bibinfo
  {author} {\bibfnamefont {J.}~\bibnamefont {Smet}}, \bibinfo {author}
  {\bibfnamefont {J.}~\bibnamefont {N\"ubler}}, \ and\ \bibinfo {author}
  {\bibfnamefont {M.}~\bibnamefont {Dolev}},\ }\bibfield  {title} {\enquote
  {\bibinfo {title} {{MBE} growth of ultra-low disorder {2DEG} with mobility
  exceeding {$35\times10^6$\,cm$^2$/Vs}},}\ }\href {\doibase
  https://doi.org/10.1016/j.jcrysgro.2008.09.151} {\bibfield  {journal}
  {\bibinfo  {journal} {J. Cryst. Growth}\ }\textbf {\bibinfo {volume} {311}},\
  \bibinfo {pages} {1658--1661} (\bibinfo {year} {2009})}\BibitemShut {NoStop}%
\bibitem [{\citenamefont {Chung}\ \emph {et~al.}(2021)\citenamefont {Chung},
  \citenamefont {{Villegas Rosales}}, \citenamefont {Baldwin}, \citenamefont
  {Madathil}, \citenamefont {West}, \citenamefont {Shayegan},\ and\
  \citenamefont {Pfeiffer}}]{Chung:2021_NM}%
  \BibitemOpen
  \bibfield  {author} {\bibinfo {author} {\bibfnamefont {Yoon~Jang}\
  \bibnamefont {Chung}}, \bibinfo {author} {\bibfnamefont {K.~A.}\ \bibnamefont
  {{Villegas Rosales}}}, \bibinfo {author} {\bibfnamefont {K.~W.}\ \bibnamefont
  {Baldwin}}, \bibinfo {author} {\bibfnamefont {P.~T.}\ \bibnamefont
  {Madathil}}, \bibinfo {author} {\bibfnamefont {K.~W.}\ \bibnamefont {West}},
  \bibinfo {author} {\bibfnamefont {M.}~\bibnamefont {Shayegan}}, \ and\
  \bibinfo {author} {\bibfnamefont {L.~N.}\ \bibnamefont {Pfeiffer}},\
  }\bibfield  {title} {\enquote {\bibinfo {title} {Ultra-high-quality
  two-dimensional electron systems},}\ }\href {\doibase
  10.1038/s41563-021-00942-3} {\bibfield  {journal} {\bibinfo  {journal} {Nat.
  Mater.}\ }\textbf {\bibinfo {volume} {20}},\ \bibinfo {pages} {632--637}
  (\bibinfo {year} {2021})}\BibitemShut {NoStop}%
\bibitem [{\citenamefont {Tsukazaki}\ \emph {et~al.}(2010)\citenamefont
  {Tsukazaki}, \citenamefont {Akasaka}, \citenamefont {Nakahara}, \citenamefont
  {Ohno}, \citenamefont {Ohno}, \citenamefont {Maryenko}, \citenamefont
  {Ohtomo},\ and\ \citenamefont {Kawasaki}}]{Tsukazaki:2010_NM}%
  \BibitemOpen
  \bibfield  {author} {\bibinfo {author} {\bibfnamefont {A.}~\bibnamefont
  {Tsukazaki}}, \bibinfo {author} {\bibfnamefont {S.}~\bibnamefont {Akasaka}},
  \bibinfo {author} {\bibfnamefont {K.}~\bibnamefont {Nakahara}}, \bibinfo
  {author} {\bibfnamefont {Y.}~\bibnamefont {Ohno}}, \bibinfo {author}
  {\bibfnamefont {H.}~\bibnamefont {Ohno}}, \bibinfo {author} {\bibfnamefont
  {D.}~\bibnamefont {Maryenko}}, \bibinfo {author} {\bibfnamefont
  {A.}~\bibnamefont {Ohtomo}}, \ and\ \bibinfo {author} {\bibfnamefont
  {M.}~\bibnamefont {Kawasaki}},\ }\bibfield  {title} {\enquote {\bibinfo
  {title} {Observation of the fractional quantum {Hall} effect in an oxide},}\
  }\href {\doibase 10.1038/NMAT2874} {\bibfield  {journal} {\bibinfo  {journal}
  {Nat. Mater.}\ }\textbf {\bibinfo {volume} {9}},\ \bibinfo {pages} {889--893}
  (\bibinfo {year} {2010})}\BibitemShut {NoStop}%
\bibitem [{\citenamefont {Piot}\ \emph {et~al.}(2010)\citenamefont {Piot},
  \citenamefont {Kunc}, \citenamefont {Potemski}, \citenamefont {Maude},
  \citenamefont {Betthausen}, \citenamefont {Vogl}, \citenamefont {Weiss},
  \citenamefont {Karczewski},\ and\ \citenamefont {Wojtowicz}}]{Piot:2010_PRB}%
  \BibitemOpen
  \bibfield  {author} {\bibinfo {author} {\bibfnamefont {B.~A.}\ \bibnamefont
  {Piot}}, \bibinfo {author} {\bibfnamefont {J.}~\bibnamefont {Kunc}}, \bibinfo
  {author} {\bibfnamefont {M.}~\bibnamefont {Potemski}}, \bibinfo {author}
  {\bibfnamefont {D.~K.}\ \bibnamefont {Maude}}, \bibinfo {author}
  {\bibfnamefont {C.}~\bibnamefont {Betthausen}}, \bibinfo {author}
  {\bibfnamefont {A.}~\bibnamefont {Vogl}}, \bibinfo {author} {\bibfnamefont
  {D.}~\bibnamefont {Weiss}}, \bibinfo {author} {\bibfnamefont
  {G.}~\bibnamefont {Karczewski}}, \ and\ \bibinfo {author} {\bibfnamefont
  {T.}~\bibnamefont {Wojtowicz}},\ }\bibfield  {title} {\enquote {\bibinfo
  {title} {Fractional quantum {Hall} effect in {CdTe}},}\ }\href {\doibase
  10.1103/PhysRevB.82.081307} {\bibfield  {journal} {\bibinfo  {journal} {Phys.
  Rev. B}\ }\textbf {\bibinfo {volume} {82}},\ \bibinfo {pages} {081307(R)}
  (\bibinfo {year} {2010})}\BibitemShut {NoStop}%
\bibitem [{\citenamefont {Betthausen}\ \emph {et~al.}(2014)\citenamefont
  {Betthausen}, \citenamefont {Giudici}, \citenamefont {Iankilevitch},
  \citenamefont {Preis}, \citenamefont {Kolkovsky}, \citenamefont {Wiater},
  \citenamefont {Karczewski}, \citenamefont {Piot}, \citenamefont {Kunc},
  \citenamefont {Potemski}, \citenamefont {Wojtowicz},\ and\ \citenamefont
  {Weiss}}]{Betthausen:2014_PRB}%
  \BibitemOpen
  \bibfield  {author} {\bibinfo {author} {\bibfnamefont {C.}~\bibnamefont
  {Betthausen}}, \bibinfo {author} {\bibfnamefont {P.}~\bibnamefont {Giudici}},
  \bibinfo {author} {\bibfnamefont {A.}~\bibnamefont {Iankilevitch}}, \bibinfo
  {author} {\bibfnamefont {C.}~\bibnamefont {Preis}}, \bibinfo {author}
  {\bibfnamefont {V.}~\bibnamefont {Kolkovsky}}, \bibinfo {author}
  {\bibfnamefont {M.}~\bibnamefont {Wiater}}, \bibinfo {author} {\bibfnamefont
  {G.}~\bibnamefont {Karczewski}}, \bibinfo {author} {\bibfnamefont {B.~A.}\
  \bibnamefont {Piot}}, \bibinfo {author} {\bibfnamefont {J.}~\bibnamefont
  {Kunc}}, \bibinfo {author} {\bibfnamefont {M.}~\bibnamefont {Potemski}},
  \bibinfo {author} {\bibfnamefont {T.}~\bibnamefont {Wojtowicz}}, \ and\
  \bibinfo {author} {\bibfnamefont {D.}~\bibnamefont {Weiss}},\ }\bibfield
  {title} {\enquote {\bibinfo {title} {Fractional quantum {Hall} effect in a
  dilute magnetic semiconductor},}\ }\href {\doibase
  10.1103/PhysRevB.90.115302} {\bibfield  {journal} {\bibinfo  {journal} {Phys.
  Rev. B}\ }\textbf {\bibinfo {volume} {90}},\ \bibinfo {pages} {115302}
  (\bibinfo {year} {2014})},\ \bibinfo {note} {also, T. Wojtowicz, private
  communication.}\BibitemShut {Stop}%
\bibitem [{\citenamefont {Pautrat}\ \emph {et~al.}(1985)\citenamefont
  {Pautrat}, \citenamefont {Francou}, \citenamefont {Magnea}, \citenamefont
  {Molva},\ and\ \citenamefont {Saminadayar}}]{Pautrat:1985_JCG}%
  \BibitemOpen
  \bibfield  {author} {\bibinfo {author} {\bibfnamefont {J.L.}\ \bibnamefont
  {Pautrat}}, \bibinfo {author} {\bibfnamefont {J.M.}\ \bibnamefont {Francou}},
  \bibinfo {author} {\bibfnamefont {N.}~\bibnamefont {Magnea}}, \bibinfo
  {author} {\bibfnamefont {E.}~\bibnamefont {Molva}}, \ and\ \bibinfo {author}
  {\bibfnamefont {K.}~\bibnamefont {Saminadayar}},\ }\bibfield  {title}
  {\enquote {\bibinfo {title} {Donors and acceptors in tellurium compounds; the
  problem of doping and self-compensation},}\ }\href {\doibase
  https://doi.org/10.1016/0022-0248(85)90143-5} {\bibfield  {journal} {\bibinfo
   {journal} {J. Crys. Growth}\ }\textbf {\bibinfo {volume} {72}},\ \bibinfo
  {pages} {194--204} (\bibinfo {year} {1985})}\BibitemShut {NoStop}%
\bibitem [{\citenamefont {Fraizzoli}\ and\ \citenamefont
  {Pasquarello}(1991)}]{Fraizzoli:1991_PRB}%
  \BibitemOpen
  \bibfield  {author} {\bibinfo {author} {\bibfnamefont {S.}~\bibnamefont
  {Fraizzoli}}\ and\ \bibinfo {author} {\bibfnamefont {A.}~\bibnamefont
  {Pasquarello}},\ }\bibfield  {title} {\enquote {\bibinfo {title} {Infrared
  transitions between shallow acceptor states in {GaAs-Ga$_{1-x}$Al$_x$As}
  quantum wells},}\ }\href {\doibase 10.1103/PhysRevB.44.1118} {\bibfield
  {journal} {\bibinfo  {journal} {Phys. Rev. B}\ }\textbf {\bibinfo {volume}
  {44}},\ \bibinfo {pages} {1118--1127} (\bibinfo {year} {1991})}\BibitemShut
  {NoStop}%
\bibitem [{\citenamefont {Kozlov}\ \emph {et~al.}(2019)\citenamefont {Kozlov},
  \citenamefont {Rumyantsev},\ and\ \citenamefont {Morozov}}]{Kozlov:2019_S}%
  \BibitemOpen
  \bibfield  {author} {\bibinfo {author} {\bibfnamefont {D.~V.}\ \bibnamefont
  {Kozlov}}, \bibinfo {author} {\bibfnamefont {V.~V.}\ \bibnamefont
  {Rumyantsev}}, \ and\ \bibinfo {author} {\bibfnamefont {S.~V.}\ \bibnamefont
  {Morozov}},\ }\bibfield  {title} {\enquote {\bibinfo {title} {Spectra of
  double acceptors in layers of barriers and quantum wells of {HgTe/CdHgTe}
  heterostructures},}\ }\href {\doibase 10.1134/S1063782619090100} {\bibfield
  {journal} {\bibinfo  {journal} {Semiconductors}\ }\textbf {\bibinfo {volume}
  {53}},\ \bibinfo {pages} {1198--1202} (\bibinfo {year} {2019})}\BibitemShut
  {NoStop}%
\bibitem [{\citenamefont {Novik}\ \emph {et~al.}(2005)\citenamefont {Novik},
  \citenamefont {Pfeuffer-Jeschke}, \citenamefont {Jungwirth}, \citenamefont
  {Latussek}, \citenamefont {Becker}, \citenamefont {Landwehr}, \citenamefont
  {Buhmann},\ and\ \citenamefont {Molenkamp}}]{Novik:2005_PRB}%
  \BibitemOpen
  \bibfield  {author} {\bibinfo {author} {\bibfnamefont {E.~G.}\ \bibnamefont
  {Novik}}, \bibinfo {author} {\bibfnamefont {A.}~\bibnamefont
  {Pfeuffer-Jeschke}}, \bibinfo {author} {\bibfnamefont {T.}~\bibnamefont
  {Jungwirth}}, \bibinfo {author} {\bibfnamefont {V.}~\bibnamefont {Latussek}},
  \bibinfo {author} {\bibfnamefont {C.~R.}\ \bibnamefont {Becker}}, \bibinfo
  {author} {\bibfnamefont {G.}~\bibnamefont {Landwehr}}, \bibinfo {author}
  {\bibfnamefont {H.}~\bibnamefont {Buhmann}}, \ and\ \bibinfo {author}
  {\bibfnamefont {L.~W.}\ \bibnamefont {Molenkamp}},\ }\bibfield  {title}
  {\enquote {\bibinfo {title} {Band structure of semimagnetic
  {Hg$_{1-y}$Mn$_y$Te} quantum wells},}\ }\href {\doibase
  10.1103/PhysRevB.72.035321} {\bibfield  {journal} {\bibinfo  {journal} {Phys.
  Rev. B}\ }\textbf {\bibinfo {volume} {72}},\ \bibinfo {pages} {035321}
  (\bibinfo {year} {2005})}\BibitemShut {NoStop}%
\bibitem [{\citenamefont {Sercel}\ and\ \citenamefont
  {Vahala}(1990)}]{Sercel:1990_PRB}%
  \BibitemOpen
  \bibfield  {author} {\bibinfo {author} {\bibfnamefont {P.~C.}\ \bibnamefont
  {Sercel}}\ and\ \bibinfo {author} {\bibfnamefont {K.~J.}\ \bibnamefont
  {Vahala}},\ }\bibfield  {title} {\enquote {\bibinfo {title} {Analytical
  formalism for determining quantum-wire and quantum-dot band structure in the
  multiband envelope-function approximation},}\ }\href {\doibase
  10.1103/PhysRevB.42.3690} {\bibfield  {journal} {\bibinfo  {journal} {Phys.
  Rev. B}\ }\textbf {\bibinfo {volume} {42}},\ \bibinfo {pages} {3690--3710}
  (\bibinfo {year} {1990})}\BibitemShut {NoStop}%
\bibitem [{\citenamefont {K\"onig}\ \emph {et~al.}(2013)\citenamefont
  {K\"onig}, \citenamefont {Baenninger}, \citenamefont {Garcia}, \citenamefont
  {Harjee}, \citenamefont {Pruitt}, \citenamefont {Ames}, \citenamefont
  {Leubner}, \citenamefont {Br\"une}, \citenamefont {Buhmann}, \citenamefont
  {Molenkamp},\ and\ \citenamefont {Goldhaber-Gordon}}]{Konig:2013_PRX}%
  \BibitemOpen
  \bibfield  {author} {\bibinfo {author} {\bibfnamefont {M.}~\bibnamefont
  {K\"onig}}, \bibinfo {author} {\bibfnamefont {M.}~\bibnamefont {Baenninger}},
  \bibinfo {author} {\bibfnamefont {A.~G.~F.}\ \bibnamefont {Garcia}}, \bibinfo
  {author} {\bibfnamefont {N.}~\bibnamefont {Harjee}}, \bibinfo {author}
  {\bibfnamefont {B.~L.}\ \bibnamefont {Pruitt}}, \bibinfo {author}
  {\bibfnamefont {C.}~\bibnamefont {Ames}}, \bibinfo {author} {\bibfnamefont
  {P.}~\bibnamefont {Leubner}}, \bibinfo {author} {\bibfnamefont
  {C.}~\bibnamefont {Br\"une}}, \bibinfo {author} {\bibfnamefont
  {H.}~\bibnamefont {Buhmann}}, \bibinfo {author} {\bibfnamefont {L.~W.}\
  \bibnamefont {Molenkamp}}, \ and\ \bibinfo {author} {\bibfnamefont
  {D.}~\bibnamefont {Goldhaber-Gordon}},\ }\bibfield  {title} {\enquote
  {\bibinfo {title} {Spatially resolved study of backscattering in the quantum
  spin {Hall} state},}\ }\href {\doibase 10.1103/PhysRevX.3.021003} {\bibfield
  {journal} {\bibinfo  {journal} {Phys. Rev. X}\ }\textbf {\bibinfo {volume}
  {3}},\ \bibinfo {pages} {021003} (\bibinfo {year} {2013})}\BibitemShut
  {NoStop}%
\bibitem [{\citenamefont {Szlenk}(1979)}]{Szlenk:1979_pssb_b}%
  \BibitemOpen
  \bibfield  {author} {\bibinfo {author} {\bibfnamefont {K.}~\bibnamefont
  {Szlenk}},\ }\bibfield  {title} {\enquote {\bibinfo {title} {Temperature
  dependence of electron concentration in intrinsic-like {HgTe}},}\ }\href
  {\doibase 10.1002/pssb.2220950214} {\bibfield  {journal} {\bibinfo  {journal}
  {phys. stat. sol. (b)}\ }\textbf {\bibinfo {volume} {95}},\ \bibinfo {pages}
  {445--452} (\bibinfo {year} {1979})}\BibitemShut {NoStop}%
\bibitem [{\citenamefont {Sawicki}\ \emph {et~al.}(1983)\citenamefont
  {Sawicki}, \citenamefont {Dietl}, \citenamefont {Plesiewicz}, \citenamefont
  {S{\c{e}}kowski}, \citenamefont {\'Sniadower}, \citenamefont {Baj},\ and\
  \citenamefont {Dmowski}}]{Sawicki:1983_Pr}%
  \BibitemOpen
  \bibfield  {author} {\bibinfo {author} {\bibfnamefont {M.}~\bibnamefont
  {Sawicki}}, \bibinfo {author} {\bibfnamefont {T.}~\bibnamefont {Dietl}},
  \bibinfo {author} {\bibfnamefont {W.}~\bibnamefont {Plesiewicz}}, \bibinfo
  {author} {\bibfnamefont {P.}~\bibnamefont {S{\c{e}}kowski}}, \bibinfo
  {author} {\bibfnamefont {L.}~\bibnamefont {\'Sniadower}}, \bibinfo {author}
  {\bibfnamefont {M.}~\bibnamefont {Baj}}, \ and\ \bibinfo {author}
  {\bibfnamefont {L.}~\bibnamefont {Dmowski}},\ }in\ \href {\doibase
  10.1007/3-540-11996-5_55} {\emph {\bibinfo {booktitle} {Application of High
  Magnetic Fields in Physics of Semiconductors}}},\ \bibinfo {editor} {edited
  by\ \bibinfo {editor} {\bibfnamefont {G.}~\bibnamefont {Landwehr}}}\
  (\bibinfo  {publisher} {Springer, Berlin},\ \bibinfo {year} {1983})\ \bibinfo
  {note} {{pp. 382-385}}\BibitemShut {NoStop}%
\bibitem [{\citenamefont {Dubowski}\ \emph {et~al.}(1981)\citenamefont
  {Dubowski}, \citenamefont {Dietl}, \citenamefont {Szyma{\'n}ska},\ and\
  \citenamefont {Ga{\l}{\c{a}}zka}}]{Dubowski:1981_JPCS}%
  \BibitemOpen
  \bibfield  {author} {\bibinfo {author} {\bibfnamefont {J.J.}\ \bibnamefont
  {Dubowski}}, \bibinfo {author} {\bibfnamefont {T.}~\bibnamefont {Dietl}},
  \bibinfo {author} {\bibfnamefont {W.}~\bibnamefont {Szyma{\'n}ska}}, \ and\
  \bibinfo {author} {\bibfnamefont {R.R.}\ \bibnamefont {Ga{\l}{\c{a}}zka}},\
  }\bibfield  {title} {\enquote {\bibinfo {title} {Electron scattering in
  {Cd$_{x}$Hg$_{1-x}$Te}},}\ }\href {\doibase 10.1016/0022-3697(81)90042-1}
  {\bibfield  {journal} {\bibinfo  {journal} {J. Phys. Chem. Solids}\ }\textbf
  {\bibinfo {volume} {42}},\ \bibinfo {pages} {351--362} (\bibinfo {year}
  {1981})}\BibitemShut {NoStop}%
\bibitem [{\citenamefont {Bir}(1974)}]{Bir:1974_B}%
  \BibitemOpen
  \bibfield  {author} {\bibinfo {author} {\bibfnamefont {G.~E.}\ \bibnamefont
  {Bir}, \bibfnamefont {G.~L.~Pikus}},\ }\href@noop {} {\emph {\bibinfo {title}
  {Symmetry and strain-induced effects in semiconductors}}}\ (\bibinfo
  {publisher} {John Wiley \& Sons, New York},\ \bibinfo {year}
  {1974})\BibitemShut {NoStop}%
\bibitem [{\citenamefont {Myers}\ \emph {et~al.}(2005)\citenamefont {Myers},
  \citenamefont {Poggio}, \citenamefont {Stern}, \citenamefont {Gossard},\ and\
  \citenamefont {Awschalom}}]{Myers:2005_PRL}%
  \BibitemOpen
  \bibfield  {author} {\bibinfo {author} {\bibfnamefont {R.~C.}\ \bibnamefont
  {Myers}}, \bibinfo {author} {\bibfnamefont {M.}~\bibnamefont {Poggio}},
  \bibinfo {author} {\bibfnamefont {N.~P.}\ \bibnamefont {Stern}}, \bibinfo
  {author} {\bibfnamefont {A.~C.}\ \bibnamefont {Gossard}}, \ and\ \bibinfo
  {author} {\bibfnamefont {D.~D.}\ \bibnamefont {Awschalom}},\ }\bibfield
  {title} {\enquote {\bibinfo {title} {Antiferromagnetic $s$-$d$ exchange
  coupling in {GaMnAs}},}\ }\href {\doibase 10.1103/PhysRevLett.95.017204}
  {\bibfield  {journal} {\bibinfo  {journal} {Phys. Rev. Lett.}\ }\textbf
  {\bibinfo {volume} {95}},\ \bibinfo {pages} {017204} (\bibinfo {year}
  {2005})}\BibitemShut {NoStop}%
\bibitem [{\citenamefont {Altshuler}\ \emph {et~al.}(2013)\citenamefont
  {Altshuler}, \citenamefont {Aleiner},\ and\ \citenamefont
  {Yudson}}]{Altshuler:2013_PRL}%
  \BibitemOpen
  \bibfield  {author} {\bibinfo {author} {\bibfnamefont {B.~L.}\ \bibnamefont
  {Altshuler}}, \bibinfo {author} {\bibfnamefont {I.~L.}\ \bibnamefont
  {Aleiner}}, \ and\ \bibinfo {author} {\bibfnamefont {V.~I.}\ \bibnamefont
  {Yudson}},\ }\bibfield  {title} {\enquote {\bibinfo {title} {Localization at
  the edge of a {2D} topological insulator by {Kondo} impurities with random
  anisotropies},}\ }\href {\doibase 10.1103/PhysRevLett.111.086401} {\bibfield
  {journal} {\bibinfo  {journal} {Phys. Rev. Lett.}\ }\textbf {\bibinfo
  {volume} {111}},\ \bibinfo {pages} {086401} (\bibinfo {year}
  {2013})}\BibitemShut {NoStop}%
\bibitem [{\citenamefont {Eriksson}(2013)}]{Eriksson:2013_PRB}%
  \BibitemOpen
  \bibfield  {author} {\bibinfo {author} {\bibfnamefont {E.}~\bibnamefont
  {Eriksson}},\ }\bibfield  {title} {\enquote {\bibinfo {title} {Spin-orbit
  interactions in a helical {Luttinger} liquid with a {Kondo} impurity},}\
  }\href {\doibase 10.1103/PhysRevB.87.235414} {\bibfield  {journal} {\bibinfo
  {journal} {Phys. Rev. B}\ }\textbf {\bibinfo {volume} {87}},\ \bibinfo
  {pages} {235414} (\bibinfo {year} {2013})}\BibitemShut {NoStop}%
\bibitem [{\citenamefont {Kimme}\ \emph {et~al.}(2016)\citenamefont {Kimme},
  \citenamefont {Rosenow},\ and\ \citenamefont {Brataas}}]{Kimme:2016_PRB}%
  \BibitemOpen
  \bibfield  {author} {\bibinfo {author} {\bibfnamefont {L.}~\bibnamefont
  {Kimme}}, \bibinfo {author} {\bibfnamefont {B.}~\bibnamefont {Rosenow}}, \
  and\ \bibinfo {author} {\bibfnamefont {A.}~\bibnamefont {Brataas}},\
  }\bibfield  {title} {\enquote {\bibinfo {title} {Backscattering in helical
  edge states from a magnetic impurity and {Rashba} disorder},}\ }\href
  {\doibase 10.1103/PhysRevB.93.081301} {\bibfield  {journal} {\bibinfo
  {journal} {Phys. Rev. B}\ }\textbf {\bibinfo {volume} {93}},\ \bibinfo
  {pages} {081301(R)} (\bibinfo {year} {2016})}\BibitemShut {NoStop}%
\bibitem [{\citenamefont {Majewicz}(2019)}]{Majewicz:2019_PhD}%
  \BibitemOpen
  \bibfield  {author} {\bibinfo {author} {\bibfnamefont {M.~M.}\ \bibnamefont
  {Majewicz}},\ }\emph {\bibinfo {title} {Nanostructure fabrication and
  electron transport studies in two-dimensional topological insulators (in
  Polish)}},\ \href@noop {} {Ph.D. thesis},\ \bibinfo  {school} {Insitute of
  Physics, Polish Academy of Sciences} (\bibinfo {year} {2019}),\ \bibinfo
  {note} {unpublished}\BibitemShut {NoStop}%
\bibitem [{\citenamefont {V\"ayrynen}\ \emph {et~al.}(2016)\citenamefont
  {V\"ayrynen}, \citenamefont {Geissler},\ and\ \citenamefont
  {Glazman}}]{Vayrynen:2016_PRB}%
  \BibitemOpen
  \bibfield  {author} {\bibinfo {author} {\bibfnamefont {J.~I.}\ \bibnamefont
  {V\"ayrynen}}, \bibinfo {author} {\bibfnamefont {F.}~\bibnamefont
  {Geissler}}, \ and\ \bibinfo {author} {\bibfnamefont {L.~I.}\ \bibnamefont
  {Glazman}},\ }\bibfield  {title} {\enquote {\bibinfo {title} {Magnetic
  moments in a helical edge can make weak correlations seem strong},}\ }\href
  {\doibase 10.1103/PhysRevB.93.241301} {\bibfield  {journal} {\bibinfo
  {journal} {Phys. Rev. B}\ }\textbf {\bibinfo {volume} {93}},\ \bibinfo
  {pages} {241301(R)} (\bibinfo {year} {2016})}\BibitemShut {NoStop}%
\bibitem [{\citenamefont {Filippone}\ \emph {et~al.}(2018)\citenamefont
  {Filippone}, \citenamefont {Moca}, \citenamefont {Weichselbaum},
  \citenamefont {{von Delft}},\ and\ \citenamefont
  {Mora}}]{Filippone:2018_PRB}%
  \BibitemOpen
  \bibfield  {author} {\bibinfo {author} {\bibfnamefont {M.}~\bibnamefont
  {Filippone}}, \bibinfo {author} {\bibfnamefont {C.~P.}\ \bibnamefont {Moca}},
  \bibinfo {author} {\bibfnamefont {A.}~\bibnamefont {Weichselbaum}}, \bibinfo
  {author} {\bibfnamefont {J.}~\bibnamefont {{von Delft}}}, \ and\ \bibinfo
  {author} {\bibfnamefont {C.}~\bibnamefont {Mora}},\ }\bibfield  {title}
  {\enquote {\bibinfo {title} {At which magnetic field, exactly, does the
  {Kondo} resonance begin to split? {A Fermi} liquid description of the
  low-energy properties of the {Anderson} model},}\ }\href {\doibase
  10.1103/PhysRevB.98.075404} {\bibfield  {journal} {\bibinfo  {journal} {Phys.
  Rev. B}\ }\textbf {\bibinfo {volume} {98}},\ \bibinfo {pages} {075404}
  (\bibinfo {year} {2018})}\BibitemShut {NoStop}%
\bibitem [{\citenamefont {Chatratin}\ \emph {et~al.}(2023)\citenamefont
  {Chatratin}, \citenamefont {Dou}, \citenamefont {Wei},\ and\ \citenamefont
  {Janotti}}]{Chatratin:2023_JPCL}%
  \BibitemOpen
  \bibfield  {author} {\bibinfo {author} {\bibfnamefont {I.}~\bibnamefont
  {Chatratin}}, \bibinfo {author} {\bibfnamefont {Baoying}\ \bibnamefont
  {Dou}}, \bibinfo {author} {\bibfnamefont {Su-Huai}\ \bibnamefont {Wei}}, \
  and\ \bibinfo {author} {\bibfnamefont {A.}~\bibnamefont {Janotti}},\
  }\bibfield  {title} {\enquote {\bibinfo {title} {Doping limits of
  {Phosphorus, Arsenic, and Antimony} in {CdTe}},}\ }\href {\doibase
  10.1021/acs.jpclett.2c03233} {\bibfield  {journal} {\bibinfo  {journal} {J.
  Phys. Chemi. Lett.}\ }\textbf {\bibinfo {volume} {14}},\ \bibinfo {pages}
  {273--278} (\bibinfo {year} {2023})}\BibitemShut {NoStop}%
\bibitem [{\citenamefont {Suski}\ \emph {et~al.}(1990)\citenamefont {Suski},
  \citenamefont {Wi{\'s}niewski}, \citenamefont {.Litwin-Staszewska},
  \citenamefont {Kossut}, \citenamefont {Wilamowski}, \citenamefont {Dietl},
  \citenamefont {Swiatek}, \citenamefont {Ploog},\ and\ \citenamefont
  {Knecht}}]{Suski:1990_SST}%
  \BibitemOpen
  \bibfield  {author} {\bibinfo {author} {\bibfnamefont {T.}~\bibnamefont
  {Suski}}, \bibinfo {author} {\bibfnamefont {P.}~\bibnamefont
  {Wi{\'s}niewski}}, \bibinfo {author} {\bibfnamefont {E}~\bibnamefont
  {.Litwin-Staszewska}}, \bibinfo {author} {\bibfnamefont {J.}~\bibnamefont
  {Kossut}}, \bibinfo {author} {\bibfnamefont {Z.}~\bibnamefont {Wilamowski}},
  \bibinfo {author} {\bibfnamefont {T.}~\bibnamefont {Dietl}}, \bibinfo
  {author} {\bibfnamefont {K.}~\bibnamefont {Swiatek}}, \bibinfo {author}
  {\bibfnamefont {K.}~\bibnamefont {Ploog}}, \ and\ \bibinfo {author}
  {\bibfnamefont {J.}~\bibnamefont {Knecht}},\ }\bibfield  {title} {\enquote
  {\bibinfo {title} {Pressure dependence of electron concentration and mobility
  in {GaAs}:{Si}-effects of on-site and inter-site interactions within a system
  of {DX} centres},}\ }\href {\doibase 10.1088/0268-1242/5/3/013} {\bibfield
  {journal} {\bibinfo  {journal} {Semicond. Sci. Techn.}\ }\textbf {\bibinfo
  {volume} {5}},\ \bibinfo {pages} {261--264} (\bibinfo {year}
  {1990})}\BibitemShut {NoStop}%
\bibitem [{\citenamefont {Lunde}\ and\ \citenamefont
  {Platero}(2012)}]{Lunde:2012_PRB}%
  \BibitemOpen
  \bibfield  {author} {\bibinfo {author} {\bibfnamefont {A.~M.}\ \bibnamefont
  {Lunde}}\ and\ \bibinfo {author} {\bibfnamefont {G.}~\bibnamefont
  {Platero}},\ }\bibfield  {title} {\enquote {\bibinfo {title} {Helical edge
  states coupled to a spin bath: Current-induced magnetization},}\ }\href
  {\doibase 10.1103/PhysRevB.86.035112} {\bibfield  {journal} {\bibinfo
  {journal} {Phys. Rev. B}\ }\textbf {\bibinfo {volume} {86}},\ \bibinfo
  {pages} {035112} (\bibinfo {year} {2012})}\BibitemShut {NoStop}%
\bibitem [{\citenamefont {{Del Maestro}}\ \emph {et~al.}(2013)\citenamefont
  {{Del Maestro}}, \citenamefont {Hyart},\ and\ \citenamefont
  {Rosenow}}]{Del_Maestro:2013_PRB}%
  \BibitemOpen
  \bibfield  {author} {\bibinfo {author} {\bibfnamefont {A.}~\bibnamefont {{Del
  Maestro}}}, \bibinfo {author} {\bibfnamefont {T.}~\bibnamefont {Hyart}}, \
  and\ \bibinfo {author} {\bibfnamefont {B.}~\bibnamefont {Rosenow}},\
  }\bibfield  {title} {\enquote {\bibinfo {title} {Backscattering between
  helical edge states via dynamic nuclear polarization},}\ }\href {\doibase
  10.1103/PhysRevB.87.165440} {\bibfield  {journal} {\bibinfo  {journal} {Phys.
  Rev. B}\ }\textbf {\bibinfo {volume} {87}},\ \bibinfo {pages} {165440}
  (\bibinfo {year} {2013})}\BibitemShut {NoStop}%
\end{thebibliography}

%

\end{document}